\documentclass[final,3p,times,twocolumn]{elsarticle}

\usepackage{amssymb}

\usepackage{lineno}

\usepackage{caption}
\usepackage{mathabx}
\usepackage{booktabs}
\usepackage{multirow}
\usepackage{siunitx}
\biboptions{authoryear}
\usepackage{color}
\usepackage[normalem]{ulem}         
\usepackage{pdflscape}
\usepackage{afterpage}

\makeatletter
\renewcommand{\fnum@figure}{Fig. \thefigure}
\makeatother

\license{{\textcopyright} 2018. This manuscript version is made available under the CC-BY-NC-ND 4.0 license}
\urllicense{http://creativecommons.org/licenses/by-nc-nd/4.0/}
\journal{Icarus 313, 61$-$78 (2018)}
\doi{DOI: 10.1016/j.icarus.2018.05.005}

\begin{document}

\begin{frontmatter}



\title{A new ab initio equation of state of hcp-Fe and its implication on the interior structure and mass-radius relations of rocky super-Earths}


\author[{UvA,VU}]{Kaustubh Hakim}
\author[ROB]{Attilio Rivoldini}
\author[{ROB,KUL}]{Tim Van Hoolst}
\author[{DEE,CMM}]{Stefaan Cottenier}
\author[CMM]{Jan Jaeken}
\author[Bay]{Thomas Chust}
\author[Bay]{Gerd Steinle-Neumann}

\address[UvA]{Anton Pannekoek Institute for Astronomy, University of Amsterdam, Science Park 904,
1098 XH Amsterdam, The Netherlands}
\address[VU]{Department of Earth Sciences, Vrije Universiteit, De Boelelaan 1085, 
1081 HV Amsterdam, The Netherlands}
\address[ROB]{Royal Observatory of Belgium, Avenue Circulaire 3, 
1180 Brussels, Belgium}
\address[KUL]{Institute for Astronomy, K.U.Leuven, Celestijnenlaan 200D,
3001 Leuven, Belgium}
\address[DEE]{Department of Electrical Energy, Metals, Mechanical Constructions and Systems, Ghent University,  Tech Lane Ghent Science Park $-$ Campus A, 
Technologiepark 903,
9052 Zwijnaarde, Belgium}
\address[CMM]{Center for Molecular Modeling, Ghent University, Tech Lane Ghent Science Park $-$ Campus A, 
Technologiepark 903,
9052 Zwijnaarde, Belgium}
\address[Bay]{Bayerisches Geoinstitut, Universit\"{a}t Bayreuth, 95440 Bayreuth, Germany}


\begin{abstract}

More than a third of all exoplanets can be classified as super-Earths based on radius (1$-$2~$\mathrm{R_{\Earth}}$) and mass ({\textless}~10~$\mathrm{M_{\Earth}}$). Here we model mass-radius relations based on silicate mantle and iron core equations of state to infer to first order the structure and composition range of rocky super-Earths assuming insignificant gas envelopes. As their core pressures exceed those in the Earth by an order of magnitude, significant extrapolations of equations of state for iron are required. We develop a new equation of state of hexagonal close packed (hcp) iron for super-Earth conditions (SEOS) based on density functional theory results for pressures up to 137~TPa. A comparison of SEOS and extrapolated equations of state for iron from the literature reveals differences in density of up to 4\% at 1~TPa and up to 20\% at 10~TPa. Such density differences significantly affect mass-radius relations. On mass, the effect is as large as 10\% for Earth-like super-Earths (core radius fraction of 0.5) and 20\% for Mercury-like super-Earths (core radius fraction of 0.8). We also quantify the effects of other modeling assumptions such as temperature and composition by considering extreme cases. We find that the effect of temperature on mass ({\textless} 5\%) is smaller than that resulting from the extrapolation of the equations of state of iron and lower mantle temperatures are too low to allow for rock and iron miscibility for $R$~{\textless}~1.75~$\mathrm{R_{\Earth}}$. Our end-member cases of core and mantle compositions create a spread in mass-radius curves reaching more than 50\% in terms of mass for a given planetary radius, implying that modeling uncertainties dominate over observational uncertainties for many observed super-Earths. We illustrate these uncertainties explicitly for Kepler-36b with well-constrained mass and radius. Assuming a core composition of 0.8$\rho$ Fe (equivalent to 50 mol\% S) instead of pure Fe leads to an increase of the core radius fraction from 0.53 to 0.64. Using a mantle composition of Mg­$_{0.5}$Fe$_{0.5}$SiO$_{3}$ instead of MgSiO$_{3}$ leads to a decrease of the core radius fraction to 0.33. Effects of thermal structure and the choice of equation of state for the core material on the core radius of Kepler-36b are small but non-negligible, reaching 2\% and 5\%, respectively. 

\end{abstract}

\begin{keyword}
Super-Earths \sep Interior structure \sep Exoplanets \sep Equation of state of iron \sep Mass-radius relations


\end{keyword}

\end{frontmatter}



\section{Introduction}\label{paper1:introduction}

With the discoveries of CoRoT-7b and Kepler-10b by ESA/CNES's and NASA's space telescopes, a new class of exoplanets known as super-Earths was introduced \citep{Leger2009,Batalha2011}. \citet{Borucki2011} classified super-Earths as rocky-type exoplanets in the size range 1.25$-$2~$\mathrm{R_{\Earth}}$ without a H/He envelope. Observational studies suggest an upper cut-off radius of super-Earths at around 1.5$-$1.7~$\mathrm{R_{\Earth}}$ \citep[e.g.,][]{Weiss2014,Buchhave2014,Rogers2015}. The low occurrence rate of close-in (orbital period {\textless} 100~d) planets between 1.5$-$2~$\mathrm{R_{\Earth}}$ introduces a radius gap in this size range, suggesting two distinct classes of exoplanets, super-Earths and sub-Neptunes \citep{Fulton2017}. This radius gap further supports an upper limit on the size of super-Earths. About a third of the four thousand\footnote{http://exoplanetarchive.ipac.caltech.edu} exoplanets known to this day have a radius between 1$-$2~$\mathrm{R_{\Earth}}$ or a mass of less than 10~$\mathrm{M_{\Earth}}$ and can potentially be classified as super-Earths. 

Mass and radius, which are the only observable properties for most super-Earths, are used to infer internal structure as well as surface and internal dynamics of exoplanets \citep[e.g.,][]{Valencia2006,Seager2007,Wagner2011}. In such studies, the mass-radius relations of solid exoplanets are computed based on assumptions about structure, composition and equation of state. The equations of state used in these studies are usually obtained by fitting material properties measured or computed at pressures of the Earth ({\textless} 300~GPa). Since the central pressure of a super-Earth with 10~$\mathrm{M_{\Earth}}$ is $\sim$4~TPa \citep{Wagner2011}, the application of these equations of state require extrapolations into pressure domains where the predicted material properties are not reliable and their usage can introduce significant errors on the interior properties of super-Earths \citep{Wagner2011}. Consequently, such errors would produce uncertainties in mass-radius relations, the extent of which has not been studied previously. Due to the scarcity in experimental data and knowledge of the stability of iron at pressures beyond those in center of the Earth \citep{Grasset2009}, the choice of an appropriate equation of state of iron for modeling the interior structure of super-Earths is non-trivial. With the help of Density Functional Theory (DFT), it becomes possible to reliably determine the equation of state and stability of a material phase at such high pressures beyond experimental limits \citep[e.g.,][]{Stixrude2014}. 

In this study, we assess the importance of various sources of uncertainty related to the equation of state of iron, to unknown core and mantle compositions and temperature in mass-radius relations of super-Earths. We assume a rocky composition for super-Earths, i.e., silicate minerals in the mantle and iron alloys in the core; we ignore water, ices or atmospheres. Our composition and structural modeling framework is given in Section~\ref{paper1:method}. For mantle mineralogy, we apply a self-consistent approach to thermodynamics. In Section~\ref{paper1:abInitioEOS}, we develop a new equation of state of iron based on ab initio calculations within DFT at pressures relevant to the core of super-Earths and compare it with other equations of state from the literature. In Section~\ref{paper1:effectEOS}, we demonstrate the effects of using extrapolations of equations of state of iron on mass-radius relations of super-Earths. We quantify the effects of temperature and core and mantle composition in Section~\ref{paper1:effectOthers}. Observed super-Earths are plotted in a mass-radius diagram in order to illustrate the extent of uncertainties in determining their structure and composition. As an example, we quantify the effects of the choice of a equation of state of iron, temperature, core and mantle composition on the interior structure of Kepler-36b, a super-Earth with well-constrained mass and radius, in Section~\ref{paper1:interiorKepler36b}.

\section{Modeling of Interior Structure and Composition}\label{paper1:method}

\subsection{Structure}\label{paper1:structure}

Structural modeling of a planet requires the computation of thermoelastic properties that depend on pressure $P$ and temperature $T$. We assume a fully differentiated planetary interior having two concentric spherical shells, mantle and core, with homogeneous chemical composition. Super-Earths might not all be differentiated due to miscibility of iron and silicates at very high temperatures as suggested by \citet{Wahl2015}, and we will briefly discuss that possibility in Section \ref{paper1:effectTemp}, but focus on the interpretation of mass and radius in terms of a differentiated interior structure. The shells in our models are considered to be spherically symmetric and isotropic such that the material properties depend only on radius $r$. To compute the planetary mass we integrate the equation describing the hydrostatic equilibrium and Poisson's equation from the center to the surface (with radius $R$). The equations are written as:
\begin{equation}\label{eq:HE}
\frac{dP}{dr} = - \rho g, 
\end{equation} 
\begin{equation}\label{eq:PE}
\frac{dg}{dr} = 4 \pi G \rho - 2 \frac{g}{r}, 
\end{equation} where $g$ is gravity, $G$ is the gravitational constant and the density $\rho(P,T)$ is calculated using an equation of state as described below. The planet's mass is then: $M = R^{2} g(R)/G$.

The heat transport mechanism in the mantle is mainly controlled by viscosity, thermal conductivity and thermal expansivity \citep[e.g.,][]{Schubert2001}. Since both viscosity and conductivity increase and expansivity decreases with pressure, it is expected that the vigor of convection decreases with depth. Whether the mantle of super-Earths is convecting as a whole or only partially is still debated as relevant transport and thermoeleastic properties are not known. However, parameterized thermal \citep{Stamenkovic2012,Wagner2012} and 2d numerical convection \citep{Tackley2013} studies, using plausible assumptions about the pressure dependence of material properties, suggest that the deep mantle of large super-Earths is convecting but with a depth-increasing sluggishness, implying a super-heated lower mantle with a strong super-adiabatic thermal gradient. For our modeling we will either assume an isentrope in the whole mantle or a temperature profile that follows an isentrope in the upper mantle and a super-adiabatic gradient in the lower mantle. Since there are indications for a strong increase of the viscosity of post-perovskite (ppv) with pressure \citep{Tackley2013}, we pin the transition to the super-adiabatic temperature at the depth of the transition to ppv. The effect of these assumptions on the mass-radius relation will be discussed in Section \ref{paper1:effectTemp}. A vigorously convective layer is essentially isentropic and the gradient of the temperature is then given by 

\begin{equation}\label{eq:TG}
\frac{dT}{dP} = \frac{\gamma}{K_{S}} T, 
\end{equation} where $\gamma$ is the Gr{\"u}neisen parameter and $K_{S}$ is the adiabatic bulk modulus. For a given potential temperature, we compute the temperature in the mantle either by solving Eq. (\ref{eq:TG}) or by computing a path of constant-entropy associated with the mantle's composition directly (see below). The thickness of the lithosphere is neglected and we use the footing temperature of the isentrope (1650~K) as the surface temperature $T_{\mathrm{S}}$.

As can be seen in Fig.~\ref{fig:AdiabatMantle}, the melting temperature of $\mathrm{MgSiO_{3}}$ \citep{Belonoshko2005} is much higher than the isentropic temperature of the mantle for pure $\mathrm{MgSiO_{3}}$ at all pressures. This simple comparison suggests that the mantles of super-Earths are likely to be solid, although higher core temperatures might imply molten lower mantles. A strong viscosity increase in the lower mantle due to high pressure and the presence of ppv may inhibit convection. Less efficient cooling of the core and lower mantle may then lead to higher temperatures. Nevertheless, our calculations assume a solid mantle. 

The thermal state of the core is mainly determined by the heat flow from the core to the mantle and thus depends on the capacity of the mantle to cool. If the heat flow out of the core is larger than what can be conducted along an adiabat, the whole core is convecting; if it is smaller, the core is at least partially thermally stratified. The extent of the stratified layer is mainly controlled by internal energy sources of the core (e.g., latent heat and chemical energy) that depend on poorly constrained material properties, on the composition of the core, and on its cooling rate. For this reason, we consider two end-member cases: a core convecting as a whole and an isothermal core.

At the core-mantle boundary (CMB), we consider two extreme scenarios for core temperature. In the first case, we assume that there is no thermal boundary, corresponding to very efficient cooling of the core. In the second case, we assume that the temperature of the core at the CMB ($T_{\mathrm{CMB}}$) is equal to the melting temperature of $\mathrm{MgSiO_{3}}$ \citep[e.g.,][]{Stixrude2014}, $T_{\mathrm{CMB}} = 1831~\mathrm{K} (1+P/4.6~\mathrm{GPa})^{0.33}$ \citep{Belonoshko2005}. Such an assumption combined with an isentropic mantle leads to an extreme temperature jump of $\sim$2500$-$7000~K depending on the mantle temperature at the CMB. These values are higher than the temperature jumps of 800$-$2000~K used by \citet{Sotin2007,Valencia2007a,Wagner2011} and serve as upper bounds in our study. 

We solve for $P$ and $g$ using Eqs. (\ref{eq:HE}), (\ref{eq:PE}) and an equation of state corresponding to the composition of the core or the mantle. We apply six boundary conditions: $g(0)=0$, $P(R)=0$ and $T_{\mathrm{S}}=1650$~K, two conditions expressing the continuity of $P$ and $g$ at the CMB; $T$ at the CMB is either continuous or equal to the melting temperature of the mantle. In calculations of model super-Earths, we fix $R$ and the core radius fraction $r_{\mathrm{CMB}}$ to compute $M$. In calculations of a super-Earth with given $M$ and $R$, we compute $r_{\mathrm{CMB}}$. 

\subsection{Mantle Modeling}\label{paper1:mantleModel}

In the context of rocky exoplanets, the mantle is commonly assumed to be made up of MgO~end-member silicate minerals and the composition is usually assumed to be $\mathrm{MgSiO_{3}}$ or $\mathrm{Mg_{2}SiO_{4}}$ \citep[e.g.,][]{Seager2007,Wagner2011}. However, a mantle with only MgO and SiO$_{2}$ is unlikely to appear in nature. Mantle minerals are expected to also incorporate oxides CaO, FeO and Al$_{2}$O$_{3}$ in addition to MgO and SiO$_{2}$, the CFMAS system, which makes up for about 99 wt\% of the bulk silicate Earth \citep{Javoy2010}. Neglecting FeO, CaO and Al$_{2}$O$_{3}$ bearing minerals that could be present in significant amounts and which mostly have higher densities than the end-member minerals in the MgO$-$SiO$_{2}$ system, would create a bias in the inference of the exoplanet's interior structure from mass and radius. \citet{Valencia2006} therefore evaluated the effect of FeO on the planetary structure by adding up to 10 wt\% of FeO to $\mathrm{Mg_{2}SiO_{4}}$ in the upper mantle and to $\mathrm{MgSiO_{3}}$ in the lower mantle.

The exact oxide content of the mantle depends on several factors including stellar composition, origin of the building blocks in the protoplanetary disk, planetary migration, the degree of differentiation in the planet and redox conditions in the mantle \citep[e.g.,][]{Bond2010b,CarterBond2012b,Dorn2015,Schaefer2017a,Santos2017}. Although data are currently missing to accurately assess realistic ranges in compositions of exoplanetary mantles, significant deviations with respect to the mantle compositions of the terrestrial planets of the solar system are to be expected. Rocky exoplanets can for example be extremely depleted in Mg compared to the Earth, as has been shown by N-body simulations for the HD17051 and HD19994 planetary systems by \citet{CarterBond2012a}. Here we consider five different end-member compositions:  $\mathrm{MgSiO_{3}}$, $\mathrm{FeSiO_{3}}$, $\mathrm{Mg_{2}SiO_{4}}$, $\mathrm{Fe_{2}SiO_{4}}$ and $\mathrm{SiO_{2}}$, thereby extending the range of mantle compositions with respect to previous studies.  The FeO~end-members of olivine and bridgmanite, although much less likely than MgO~end-members, represent an upper bound on the mantle density as they are the highest density minerals in the respective solid solutions \citep{Stixrude2011}. 

\begin{figure}[!ht]
  \centering
  \medskip
  \includegraphics[width=.45\textwidth]{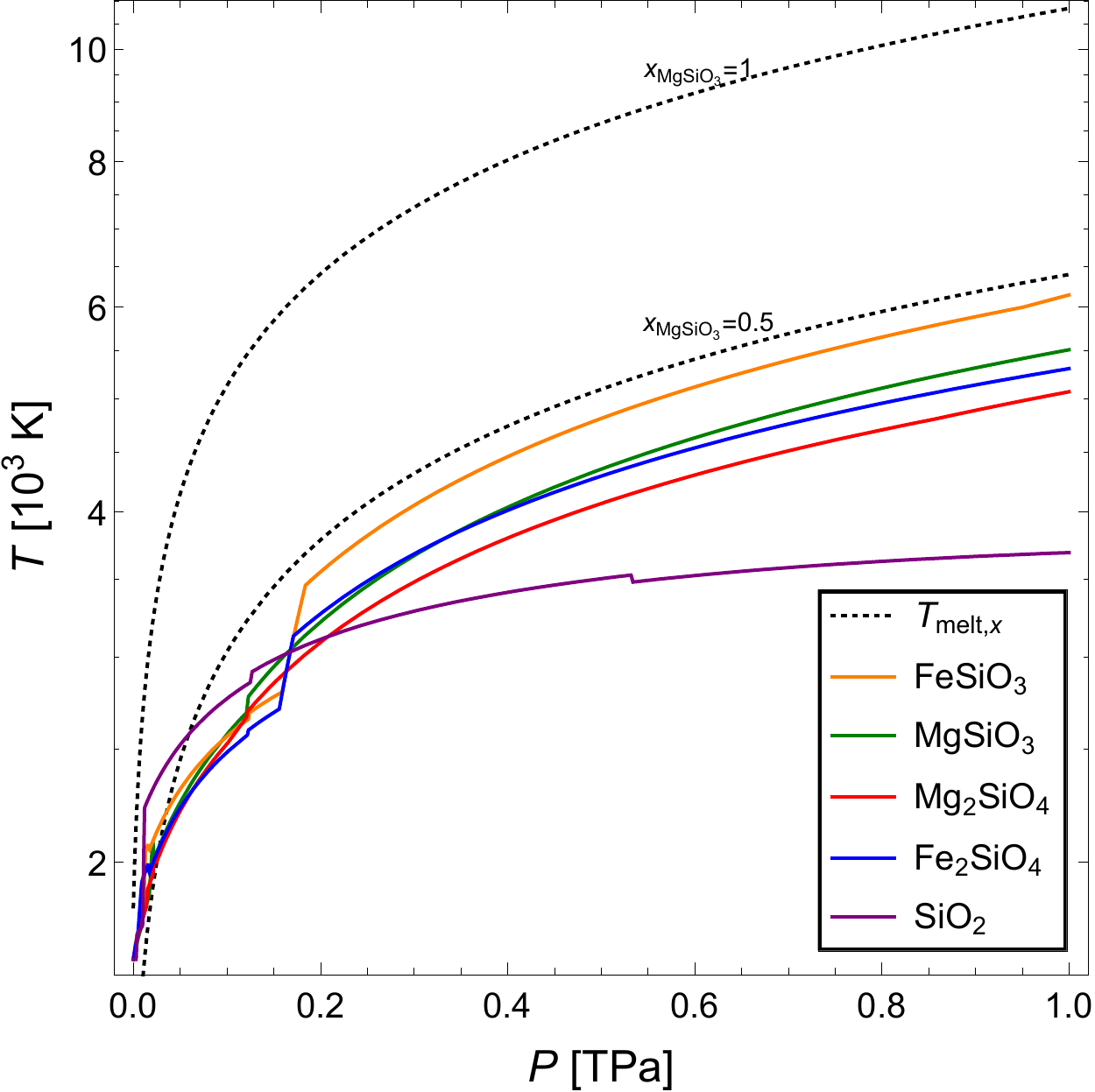}
  \caption[Isentropes of mantle minerals]{Isentropes of the mantle mineralogy computed thermodynamically self-consistently with $T_{\mathrm{S}}=1650$~K. All relevant phases of $\mathrm{Mg_{2}SiO_{4}}$, $\mathrm{Fe_{2}SiO_{4}}$, $\mathrm{MgSiO_{3}}$, $\mathrm{FeSiO_{3}}$ and $\mathrm{SiO_{2}}$ are taken into account. Dotted lines represent the melting temperature of $\mathrm{MgSiO_{3}}$, $T_{\mathrm{melt}} = 1831~\mathrm{K} (1+P/4.6~\mathrm{GPa})^{0.33}$ \citep{Belonoshko2005} with the cryoscopic equation $T_{\mathrm{melt},x}=T_{\mathrm{melt}}(1- \ln{x_{\mathrm{MgSiO_{3}}}})^{-1}$ \citep{Atkins1994}, where $x_{\mathrm{MgSiO_{3}}}$ is the mole fraction of pure $\mathrm{MgSiO_{3}}$.}
  \label{fig:AdiabatMantle}
\end{figure}

Phase relations and physical properties of mantle minerals are evaluated self-consistently using a new implementation \citep{Chust2017} of the thermodynamic model of \citet{Stixrude2011} based on a Birch-Murnaghan Mie-Gr{\"u}neisen equation of state formulation. At least for density, the central property we are interested in, the model shows robust behavior beyond the $P$$-$$T$ conditions of the lowermost mantle \citep{Connolly2016}. From the database of \citet{Stixrude2011}, we use MgO, the SiO$_{2}$ phases as well as the MgO- and FeO-bearing end-members of all silicate solid solutions to compute assemblages for the compositions given above. As the database of \citet{Stixrude2011} is only assessed for conditions (and minerals) of the Earth's mantle and ignores phase transitions at higher pressures, such as the dissociation of post-perovskite (ppv) \citep[e.g.,][]{Umemoto2011,Wu2014} or the B1-B2 transition in MgO \citep[e.g.,][]{Cebulla2014}, we investigate an alternative scenario for mantle structure (Section \ref{paper1:effectMantleComp}), taking high-pressure phases and transitions from \citet{Umemoto2011} into account. 

The isentropes associated with the five model compositions and the potential temperature of 1650~K are shown in Fig.~\ref{fig:AdiabatMantle}. Isentropes show $T$-discontinuities at phase transitions with significant Clausius-Clapeyron slopes through volume changes and latent heat effects. Mostly, temperatures increase as different compositions undergo phase transitions with positive Clapeyron slopes. The perovskite$-$ppv transition in the MgSiO$_{3}$ and Mg$_{2}$SiO$_{4}$ compositions in the vicinity of 100~GPa are such an example. The melting temperature of $\mathrm{MgSiO_{3}}$ is higher than the temperatures along the isentropes of $\mathrm{MgSiO_{3}}$ and $\mathrm{Mg_{2}SiO_{4}}$, implying a solid mantle for super-Earths. The larger $T$-jumps for the FeO-bearing compositions at a pressure between 150$-$200~GPa stem from the formation of the FeSiO$_{3}$ ppv phase from a combination of SiO$_{2}$ seifertite and FeO w{\"u}stite, as the FeSiO$_{3}$ perovskite phase is predicted stable only at temperatures below 1500~K. For an SiO$_{2}$-only mantle, the stishovite-seifertite transition between 100$-$150~GPa leads to a temperature increase, but stishovite is then predicted to form again near 500~GPa with a negative Clapeyron slope and a slight temperature decrease. The re-appearance of stishovite is a numerical artifact of the database calibration, and a proper description of an SiO$_{2}$-based mantle would require an explicit consideration of the predicted high-pressure phases \citep{Tsuchiya2011,Umemoto2011}. However, based on our discussion in Section \ref{paper1:effectMantleComp}, such a consideration will not strongly affect the overall structure of super-Earths.

\subsection{Core Modeling}\label{paper1:coreModel}

Iron is ubiquitously assumed to be the major element in the core of super-Earths \citep[e.g.,][]{Seager2007,Wagner2011}. To compute the density of Fe at relevant conditions, previous studies have extrapolated different formulations of the equation of state of Fe which are strictly speaking only valid within the pressure range of their assessment (see Table \ref{tab:LiteratureEOSTable}). For instance, the Vinet formulations of \citet{Valencia2007a} and \citet{Seager2007} and the third-order Birch-Murnaghan (BM3) formulation of \citet{Dorn2015} are fitted to data of Fe up to pressures of 200$-$300~GPa but extrapolated to 5$-$20~TPa. Neither the Vinet nor BM3 equations have extrapolations that are thermodynamically consistent to infinite pressure \citep[e.g.,][]{Holzapfel1996,Stacey2005}. \citet{Seager2007} switch to a Thomas-Fermi-Dirac equation of state at 20.9~TPa that is valid at in the infinite pressure limit. \citet{Wagner2011} have used the modified Rydberg equation \citep{Stacey2005} and fitted the data from \citet{Dewaele2006}, obtained between 17 and 197~GPa. Extrapolation to several tens of TPa will result in significantly different density predictions, depending on the equation used, even when those equations approach a thermodynamically consistent limit at infinite pressure. For this reason, we calculated an equation of state of iron for super-Earths (SEOS) up to 137~TPa with ab initio methods. With this equation of state the core of Earth-mass and ultra-massive super-Earths can be modeled without using a biasing extrapolation scheme. Details are discussed in Section \ref{paper1:formulation}. 

For the Earth's core, Fe accounts for about 85 wt\% and the remainder consists of Ni and light elements. For the light elements in the Earth's core, there is no consensus on their amount and identity. A recent synthesis of different studies about the composition by \citet{Hirose2013} suggest: Si$=$6 wt\%, O$=$3 wt\% and S$=$1-2 wt\%. Also, the possibility of C as a light element in the Earth's core cannot be ruled out \citep{Fei2014}. In order to evaluate the effect of potentially important light elements in the core on mass-radius relations, we consider two approaches, reflecting the combined effects of Ni and light elements. First, we assume a decrease in density of the core with respect to pure iron by 10\%, similar to the Earth's core; we also consider a decrease in density of 20\%, i.e., we assume the core density to be either 0.9$\rho$~Fe or 0.8$\rho$~Fe. Second, in order to be able to assess the effect of the presence of specific light elements, we assume that the core is made of either of the following Fe-alloys: FeS, FeSi, Fe$_{0.95}$O, Fe$_{3}$C (see Table \ref{tab:SataHolzapfelEOS}). For details about the equations of state, see \ref{paper1:appendixFeAlloys}.

The melting curve of iron is steeper than a core isentrope at super-Earth pressures, suggesting that the core of massive super-Earths is solid if the temperature at the core-mantle boundary is below the melting temperature \citep{Morard2011}. The effect of light elements on the melting temperature of Fe at super-Earth core conditions is not known, but a maximal depression of 1500~K at 1.5~TPa has been proposed \citep{Morard2011}. The melting temperature of such an alloy is then above $13,000$~K at 1.5~TPa and thus significantly above the predicted temperature at that pressure in super-Earth cores \citep{Valencia2006,Wagner2012}, indicating a solid core. Several factors including a sluggish or non-convecting lower mantle, warm initial conditions and a high temperature jump at the CMB \citep{Stixrude2014} would allow for significantly higher core temperatures. Nevertheless, since the density deficit between the solid and the liquid state decreases with pressure \citep{Alfe2002b} and is about 1$-$2\% at the inner core boundary conditions of the Earth \citep{Alfe2002a,Laio2000,Ichikawa2014}, the use of an equation of state describing the density of solid iron has likely a negligible effect on mass-radius relations.

\subsection{Mass-Radius Relations based on Core Size}\label{paper1:massRadiusRelations}

Since mass and radius are currently the only observable parameters for most of the rocky exoplanets, mass-radius ($M$$-$$R$) relations are used to obtain a first-order estimate of their composition. Theoretical $M$$-$$R$ relations of rocky planets have been computed for isothermal planetary interiors \citep[e.g.,][]{Zapolsky1969,Seager2007} and temperature increasing with depth \citep[e.g.,][]{Valencia2006,Fortney2007,Sotin2007,Wagner2011}. Simple power-law relations are generally used for super-Earths:

\begin{equation}\label{eq:ScalingLaw}
R = a M^\beta, 
\end{equation} where $R$ and $M$ are in Earth units, $\beta$ is the scaling exponent and $a$ is the proportionality constant. $\beta$ determines how the radius increases with mass: the larger $\beta$, the bigger are planets for the same mass. For small bodies whose internal density is not affected by pressure, $\beta=1/3$. For larger planetary bodies, adiabatic self-compression leads to a slower increase of radius with mass and therefore smaller $\beta$. 

Since the core size has a direct impact on planetary mass and bulk density, we calculate $M$$-$$R$ relations for three classes of rocky super-Earths: Mercury-like, Earth-like and Moon-like by fixing core radius fractions to 0.8, 0.5 and 0.2, respectively. These values are similar to those of Mercury (0.82), the Earth (0.54) and the Moon (0.21) \citep[e.g.,][]{DziewonskiAnderson1981,Garcia2011,Hauck2013,Rivoldini2013}. Our definition of rocky super-Earth classes contrasts with most previous studies \citep[e.g.,][]{Valencia2006,Wagner2011} in which the core mass fraction and essentially the bulk elemental composition of an exoplanet class is constant but the relative core radius changes along a $M$$-$$R$ relationship. With our definition, the bulk composition of super-Earths of a given class slightly varies with mass (e.g., elemental abundance of Mg changes by \textless 2~wt\%) since the relative core radius is constant. This difference results in small differences of the order of 1$-$4\% in the $\beta$ values of the $M$$-$$R$ relations. In order to be able to make consistent comparisons with the literature, we recompute $\beta$ values for previous studies based on the same interior structure assumptions as in our work (see Table \ref{tab:BetaEOSTable}). We also compute $M$$-$$R$ relations for bare-core super-Earths since observations suggest their existence \citep{Mocquet2014}. Such a calculation provides an upper bound on the masses of super-Earths as a function of radius.

\section{A New ab Initio Equation of State of hcp-Fe}\label{paper1:abInitioEOS}

\subsection{Formulation of SEOS}\label{paper1:formulation}

Laser-heated diamond-anvil cell experiments \citep{Sakai2011} show that the crystal structure of pure Fe at the conditions of the Earth's core is hexagonal close-packed (hcp). However, high core temperatures \citep{Belonoshko2017}, the presence of nickel \citep{Dubrovinksy2007} and light elements in the core \citep{Vocadlo2003} could stabilize the body-centered cubic (bcc) phase over hcp at the Earth's inner-core conditions. The exact crystal phase of the Earth's inner core remains a matter of extensive debate, but hcp-Fe is the most plausible candidate \citep[e.g.,][]{Hirose2013}.

Ab initio crystal structure predictions at zero temperature show that hcp-Fe is more stable than bcc and fcc up to 8~TPa and between 24$-$35~TPa \citep[e.g.,][]{PickardNeeds2009b,Cottenier2011}, and this remains true up to at least $10,000$~K \citep{Stixrude2012}. Between 8$-$24~TPa, the face-centered cubic (fcc) phase is stable and above 35~TPa the bcc phase takes over. Since the expected core pressures in Earth-like super-Earths are likely below 10~TPa \citep{Wagner2011}, their cores would be dominated by hcp-Fe if they are pure~Fe and at zero temperature. However, as already mentioned, high temperature and alloying elements will likely change the stability fields of different phases. The exact phase at TPa pressures is probably not very important for mass-radius relations of super-Earths since the density difference between the crystal phases of iron is expected to be small, especially for the closed-packed phases. According to the calculations in \citet{Cottenier2011}, at 350~GPa fcc is 0.05\% less dense than hcp, while bcc is 1.2\% less dense. At higher pressures, density differences remain small: At 8~TPa, fcc and bcc are denser than hcp by less than 0.1\%. For this reason, we approximate iron in the core of super-Earths as the hcp phase.

Here we perform Density Functional Theory (DFT)-based calculations for hcp-Fe in the volume range of 0.50$-$6.79~$\mathrm{cm^{3}/mol}$, to cover the whole compression range relevant for the cores of super-Earths. For high precision, we use the all-electron augmented plane wave method with local orbitals (APW+lo) \citep{Schwarz2010,Sjostedt2000,Madsen2001,Cottenier2002} with the PBE approximation to the exchange and correlation fuctional \citep{Perdew1996}. Within the APW+lo approach, we treat the $3s$, $3p$, $3d$, $4s$, $4p$ states as valence electrons. The muffin tin radius for the Fe-atoms was set to 1.6 au, and the size of the APW+lo basis specified by $K_{\mathrm{max}}$ = 8.5/1.6~au = 5.31~au$^{-1}$. Brillouin zone integration used a mesh of $42\times42\times22 \ k$-points. Calculations were performed both for non~spin-polarized and spin-polarized charge densities, using the type II spin arrangement of \cite{Steinle2004}. At each volume, the $c/a$ ratio was optimized and we found an increase with compression from 1.5858$-$1.6165 in the range from 0$-$2.5~TPa (1.6033 at 350~GPa).

From the energy-volume relation of 181 values (see Supplement) pressure is computed as the first derivative numerically. We employ an averaged 2-point backward and forward scheme that takes into account the uneven volume sampling. We chose this strategy, as the energy-volume range sampled by our calculations is too large to be adequately fit with a single closed expression (Fig. \ref{fig:deltap_V_Holzapfel}). Without a closed expression for the equation of state, we obtain volumes at given pressure by an interpolation scheme: this is done with a method that uses a sliding-window to locally fit a finite strain expression (see \ref{paper1:appendixInterpolation} for a detailed description).

In previous studies, several methods within the framework of DFT have been applied  to calculate the equation of state of hcp-Fe, almost always with the Perdew-Burke-Ernzerhof (PBE) exchange-correlation functional or closely related ones and with a variety of numerical solvers. Our results at zero pressure (Table \ref{tab:IronEOSTable}) are in good agreement with published values, and the small differences are what can be expected when comparing different DFT codes \citep{Lejaeghere2016a}. Our density results are in excellent agreement with those of \citet{PickardNeeds2009b} over a broad pressure range \citep[cf. Fig. 1,][]{Cottenier2011} and differ by less than 0.2\% with the results of \citet{Stixrude2012} at 6 and 38.3~TPa. The accuracy of DFT (PBE) predictions for a ground state equation of state (equilibrium volume or density, bulk modulus) have been assessed by \citet{Lejaeghere2013,Lejaeghere2016b}. In particular when the bulk modulus is large, as in the present work, the uncertainty on the predicted density becomes negligibly small \citep{Lejaeghere2016b}.

Measurements of compression data at room temperature provide a reliable equation of state for iron up to almost 300~GPa \citep[e.g.,][]{Fei2016}: the third-order Birch-Murnaghan equation (BM3) used to summarize new and previous experimental data have pressure residuals that are not much larger than 2\%. A fit of the same data to a Vinet equation \mbox{\citep[e.g.,][]{Poirier2000}} results in similar residuals and the predicted bulk moduli from both equations are equivalent (see Fig.~\ref{fig:deltapKT}). The induced uncertainty on the parameters of the BM3 equation by the data uncertainties results in an error on density that is smaller than 0.5\% and that on the bulk modulus is smaller than 1\% at 300~GPa. In order to use the laboratory measurements together with the DFT results, the former have been extrapolated to 0~K \citep{Fei2016} and zero-point vibration effects have been removed \citep{Lejaeghere2014}. The equations based on prior experimental data \citep[e.g.,][]{Williams1997,Anderson2001,Dewaele2006} show deviations from the equation of state of \citet{Fei2016} and SEOS (based on the same low pressure data) that are larger than 0.5\% (Fig.~\ref{fig:EOSIronComparison}) because they have either been deduced from smaller data sets or/and the uncertainties on those data were larger. 

Our DFT results for hcp exhibit the well known pressure underestimation of about 8\% at small compressions \citep{Steinle2004,Fei2016} (Fig.~\ref{fig:LowPressureSEOS}), an effect that cannot be explained by thermal pressure alone (less than 1\%) and for which dynamical many-body effects beyond DFT need to be invoked \citep{Pourovskii2014}. This discrepancy is larger than what is usually obtained using DFT at the PBE level, including for bcc-Fe where the agreement with experiments is excellent \citep[see][]{Lejaeghere2014}. After correcting laboratory data of hcp-Fe for thermal effects, densities become comparable to DFT results near 200~GPa. At pressures higher than 300~GPa, and depending on the formulation used to summarize the experimental data, extrapolated pressures grow weaker or stronger than our DFT results with compression (see Fig.~\ref{fig:EOSIronComparison} and discussion below). Therefore, we use the experimental equation of state of hcp iron for pressures up to 234~GPa, the pressure where laboratory data and our DFT results intersect, and our DFT results for pressures above. A discontinuity in $K_{T}$ of $\sim$10\% between the experimental equation of state and our DFT results has a negligible effect on mass-radius relations and a small effect on the thermal state of the core.

In addition to the numerical determination of SEOS that we have introduced above, we include, for convenience, a closed equation of state that is valid for pressures above 234 GPa. When the Holzapfel equation \citep{Holzapfel1996} is fitted to our ab initio results in the pressure range 0.234$-$10~TPa, the volume residuals are below 1.1\%, whereas fits with BM3 or Vinet equations have residuals that are 4 to 8 times larger. The Holzapfel equation is written as 
\begin{equation}\label{eq:PressureHolzapfel}
P = P_{0} + 3 K_{T,0} x^{\frac{-5}{3}} (1-x^{\frac{1}{3}})  [1 + c_{2}x^{\frac{1}{3}}(1-x^{\frac{1}{3}})] \exp [c_{0}(1-x^{\frac{1}{3}})],
\end{equation} where $x=V/V_{0}$, $V_{0}=4.28575~\mathrm{cm^3/mol}$, $P_{0}=234.4$~GPa, $K_{T,0}=1145.7 \pm 3.6$~GPa, $c_{0}=3.19 \pm 0.08$ and $c_{2}= -2.40 \pm 0.05$. 

\subsection{Thermal Pressure}\label{paper1:thermalEOS}

Total pressure can be divided into a cold (isothermal) part (discussed above), and two temperature-dependent (thermal) parts: $P (V, T) = P (V, 0~\mathrm{K})$ $+ P_{\mathrm{harm}} (V, T) + P_{\mathrm{ae}} (V, T)$, where $P_{\mathrm{harm}}$ is the quasiharmonic thermal pressure and $P_{\mathrm{ae}}$ is the anharmonic-electronic thermal pressure. For the quasiharmonic thermal pressure, phonon calculations can be performed \citep[e.g.,][]{Vocadlo2008,Stixrude2012} or approximated by the Debye \citep[e.g.,][]{Dewaele2006,Fei2016} or Einstein models \citep[e.g.,][]{Bouchet2013}. At high temperatures in the core of super-Earths, well above the Debye temperature, the Einstein and Debye models practically coincide \citep{Poirier2000}. Following \citet{Bouchet2013}, we use the simpler Einstein model to compute $P_{\mathrm{harm}}$ and the Einstein temperature $\Theta$:

\begin{equation}\label{eq:Pharm}
P_{\mathrm{harm}}(V,T) = \frac{3 R \gamma}{V} \left[ \frac{\Theta}{2} + \frac{\Theta}{\exp{(\Theta/T)} - 1} \right],
\end{equation} 

\begin{equation}\label{eq:Theta}
\Theta = \Theta_{0} x^{-\gamma_{\infty}} \exp \left[ \frac{\gamma_{0}-\gamma_{\infty}}{b} (1-x^{b}) \right],
\end{equation} with the Gr{\"u}neisen parameter $\gamma = \gamma_{\infty} + (\gamma_{0} - \gamma_{\infty})x^{b}$, where $R$ is the universal gas constant, $b$ is a constant and $x = V/V_{0}$. The approach of \citet{Bouchet2013} reproduces the experimental $P$-$V$-$T$ equation of \citet{Dewaele2006} well that is fit to static and shock wave experiments as well as ab initio simulations. The anharmonic-electronic thermal pressure is given as

\begin{equation}\label{eq:Pae}
P_{\mathrm{ae}}(V,T) = \frac{3 R}{2 V} m a_{0} x^{m} T^{2},
\end{equation} where $m$ and $a_{0}$ are constants. 

For both the harmonic and anharmonic-electronic thermal pressure we use the parameter values estimated by \citet{Bouchet2013} (see Table \ref{tab:LiteratureEOSTable}). Since our reference volume $V_{0}$ is not the same as that of \citet{Bouchet2013}, we replace $x$ in the equations for the quasiharmonic and anharmonic-electronic thermal pressures by $x (V_{0}/V_{0,\mathrm{Bouchet}})$. With this modification, the thermal pressure is equal to that of \citet{Bouchet2013} for a given volume $V$. 

The isothermal bulk modulus $K_{T}$, the isobaric thermal expansivity $\alpha$ and the adiabatic bulk modulus $K_{S}$ are given by 

\begin{equation}\label{eq:BulkMod}
K_{T} = - V \left( \frac{\partial P(V,T)}{\partial V} \right)_{T},
\end{equation} 

\begin{equation}\label{eq:Alpha}
\alpha = \frac{1}{V} \left( \frac{\partial V}{\partial T} \right)_{P}, 
\end{equation} 

\begin{equation}\label{eq:AdiBulkMod}
K_{S} = K_{T} (1 + \alpha \gamma T).
\end{equation}

\subsection{Comparison of Extrapolations of the Equations of State of Iron}\label{paper1:comparisonEOS}

We compare previously published equations of state of Fe (see Table \ref{tab:LiteratureEOSTable}) with our equation of state for super-Earths (SEOS) in the pressure range 0.234$-$10~TPa at 0~K. The various equations determined at a finite temperature were corrected to 0~K according to the method described in the respective articles and the zero-point vibration effects have been removed. 

\begin{figure}[!ht]
  \centering
  \medskip
  \includegraphics[width=.45\textwidth]{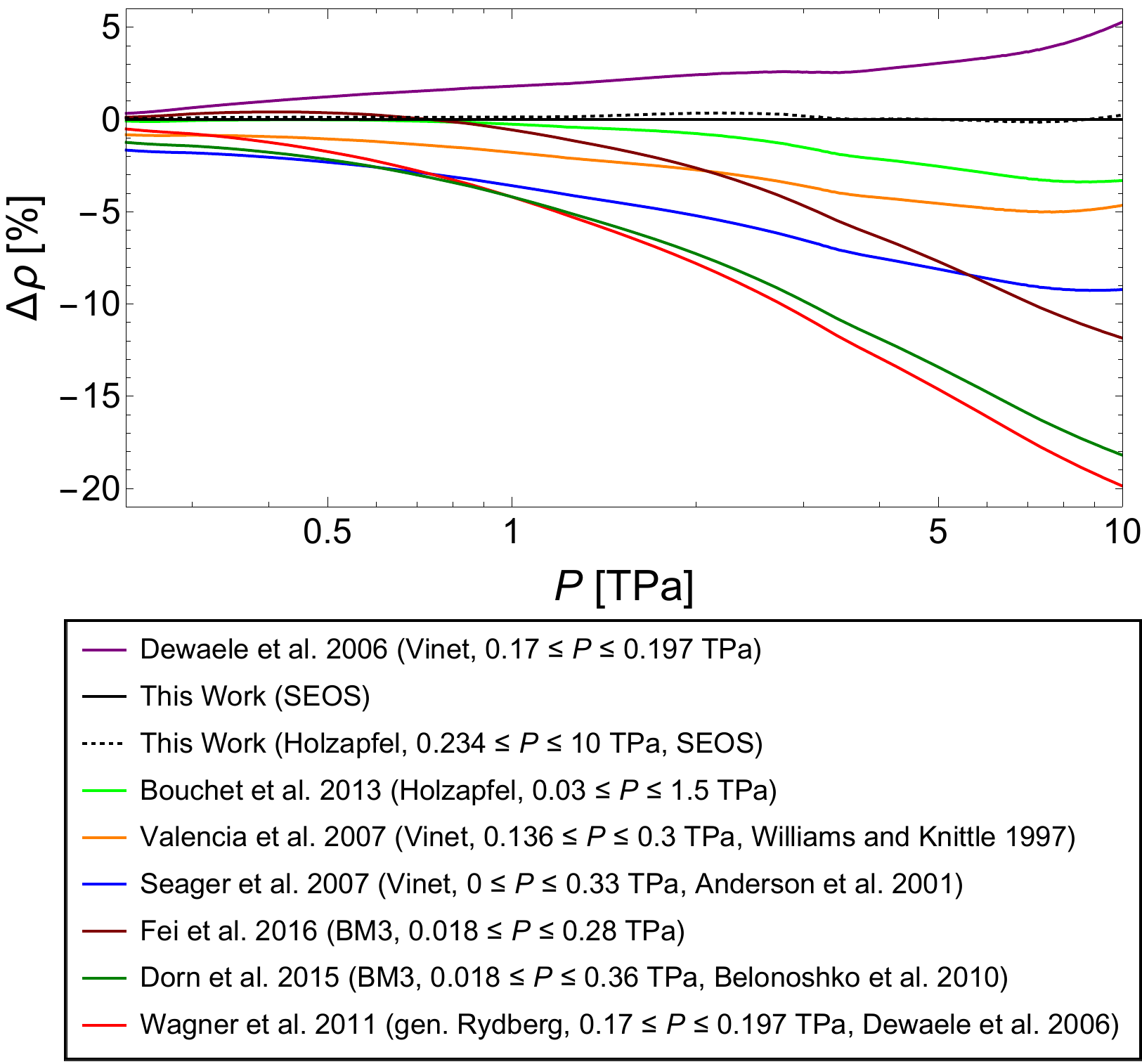}
  \caption[Comparison of the equation of state of iron]{Comparison of the relative density difference in per cent for equations of state of iron listed in Table \ref{tab:LiteratureEOSTable} and for the Holzapfel fit to SEOS up to 10~TPa relative to SEOS at 0~K in the pressure range of 0.234$-$10~TPa. These equations of state are indicated by references, the form of the equation used, the pressure range of the fit and, if applicable, the source of the data. }
  \label{fig:EOSIronComparison}
\end{figure}

The BM3, Vinet, modified Rydberg and Holzapfel equations describe compression data with similar precision below 0.3~TPa if no prior assumption on equation parameters are made (see \ref{paper1:appendixExtrapolation}), but extrapolation yields significantly different predictions at high pressures (see Fig.~\ref{fig:EOSIronComparison}). Among those equations, the modified Rydberg and Holzapfel equations show a consistent thermodynamic behavior at infinite pressure and should therefore be used if large pressure extrapolations are required. The relevance of this behavior is illustrated by the equation of state from \citet{Bouchet2013} at 10~TPa (Fig.~\ref{fig:EOSIronComparison}). Its density deviates by not much more than 3\% from SEOS. In principle, a similar precision could be obtained with the modified Rydberg equation (see \ref{paper1:appendixFitting}), but the large deviations exhibited by the equation of state from \citet{Wagner2011} at 10~TPa (Fig.~\ref{fig:EOSIronComparison}) is the result of their choice of a specific value for the derivative of the bulk modulus at infinite pressure prior to fitting the laboratory data of \citet{Dewaele2006}.

\afterpage{
}
\begin{table*}[!ht]
\caption{\label{tab:MassEOSTable} Interior structure parameters and differences in planetary mass between SEOS and other equations of state of iron for different core sizes and planetary radii. Differences below 0.05\% are not reported. Models assume an isentropic pure~Fe core, an isentropic $\mathrm{MgSiO_{3}}$ mantle, $T_{\mathrm{S}}=1650$~K and a zero CMB $T$-jump. }
\scriptsize
\begin{center}
  \begin{tabular}{lll|cccc|cccc} \hline \hline \rule[0mm]{0mm}{0mm} 
  
{Ref.} & {Form.} & {Parameter} & {Bare-core} & {Mercury-like} & {Earth-like} & {Moon-like} & {Bare-core} & {Mercury-like} & {Earth-like} & {Moon-like} \\
& & $r_{\mathrm{CMB}}$          & 1  & 0.8  & 0.5  & 0.2  & 1  & 0.8  & 0.5  & 0.2  \\
& & $R$ ($\mathrm{R_{\Earth}}$) & \multicolumn{4}{c}{1.25} & \multicolumn{4}{c}{1.75} \\[2mm]	

\hline
					
SEOS & $-$ & $P_{\mathrm{c}}$ (TPa) 	          & 4.6  & 2.5   & 0.9   & 0.3  & 37.8 & 16.8  & 4.0   & 1.2  \\
	 & & $P_{\mathrm{th,c}}$/$P_{\mathrm{c}}$     & 3\%  & 4\%   & 7\%   & 11\% & 1\%  & 2\%   & 4\%   & 6\%  \\
	 & & $P_{\mathrm{ae,c}}$/$P_{\mathrm{th,c}}$  & 14\% & 17\%  & 20\%  & 22\% & 8\%  & 12\%  & 16\%  & 19\% \\
	 & & $T_{\mathrm{c}}$ ($10^{3}$~K)            & 8.6  & 7.2   & 5.1   & 3.7  & 18.2 & 15.4  & 9.1   & 5.8  \\
& & $\rho_{\mathrm{c}}$ ($\mathrm{Mg/m^{3}}$)     & 30.4 & 24.4  & 17.7  & 13.6 & 66.1 & 48.9  & 28.8  & 19.1 \\
	 & & $M_{core}/M$                             & 100\%& 79.3\%& 31.0\%& 2.4\%& 100\%& 83.0\%& 33.7\%& 2.4\%\\
	 & & $M$ ($\mathrm{M_{\Earth}}$)              & 6.5  & 4.0   & 2.2   & 1.7  & 33.6 & 18.7  & 8.4   & 5.9  \\

\hline 
Dew & Vin & $\Delta M$ & $+$1\%  & $+$1\%   & $+$0.5\% & $-$  & $+$13\% & $+$10\% & $+$1.6\%  & $+$0.2\% \\
Bou & Hol & $\Delta M$ & $-$1\%  & $-$1\%   & $-$      & $-$  & $-$6\%  & $-$5\%  & $-$0.8\%  & $-$      \\
Val & Vin & $\Delta M$ & $-$2\%  & $-$1\%   & $-$      & $-$  & $-$7\%  & $-$7\%  & $-$1.7\%  & $-$      \\
Sea & Vin & $\Delta M$ & $-$5\%  & $-$3\%   & $-$      & $-$  & $-$18\% & $-$14\% & $-$3.1\%  & $-$      \\
Fei & BM3 & $\Delta M$ & $-$3\%  & $-$1\%   & $+$0.5\% & $-$  & $-$22\% & $-$14\% & $-$1.9\%  & $-$      \\
Dor & BM3 & $\Delta M$ & $-$13\% & $-$7\%   & $-$1.4\% & $-$  & $-$34\% & $-$23\% & $-$5.7\%  & $-$0.2\% \\
Wag & Ryd & $\Delta M$ & $-$12\% & $-$6\%   & $-$1.4\% & $-$  & $-$35\% & $-$24\% & $-$6.0\%  & $-$0.2\% \\[2mm]
\hline
  \end{tabular}
  \end{center}\caption*{Abbreviations: SEOS: Super-Earths equation of state, This Work, $P_{\mathrm{th}}$ from \citet{Bouchet2013}; Dew: \citet{Dewaele2006}; Bou: \citet{Bouchet2013}; Val: \citet{Valencia2007a}, data from \citet{Williams1997}; Sea: \citet{Seager2007}, data from \citet{Anderson2001}; Fei: \citet{Fei2016}; Dor: \citet{Dorn2015}, data from \citet{Belonoshko2010}; \citet{Wagner2011}, data from \citet{Dewaele2006}; Vin: Vinet; BM3: third-order Birch-Murnaghan; Hol: Holzapfel; Ryd: modified Rydberg. }
\end{table*}

Extrapolations with the Vinet formulation depart by more than 5\% from SEOS at 10~TPa (Fig.~\ref{fig:EOSIronComparison}) and deviations increase further with pressure. The large variations between the equations of state by \citet{Dewaele2006}, \citet{Seager2007} and \citet{Valencia2007a} at high pressure are the result of the different data sets used (see Table \ref{tab:LiteratureEOSTable}). Among the equations considered, the BM3 equation is the least well suited for large pressure extrapolation (see \ref{paper1:appendixExtrapolation}) and should be avoided. Note, however, that extrapolation of the equation of state up of \citet{Fei2016} to 1~TPa results in a density less than 0.5\% different from that of SEOS. This is a consequence of the extensive and precise data set used to fit the BM3 equation.

\section{Effect of the Iron Equation of State on M-R Relations}\label{paper1:effectEOS}

\begin{figure}[!ht]
  \centering
  \medskip
  \includegraphics[width=.45\textwidth]{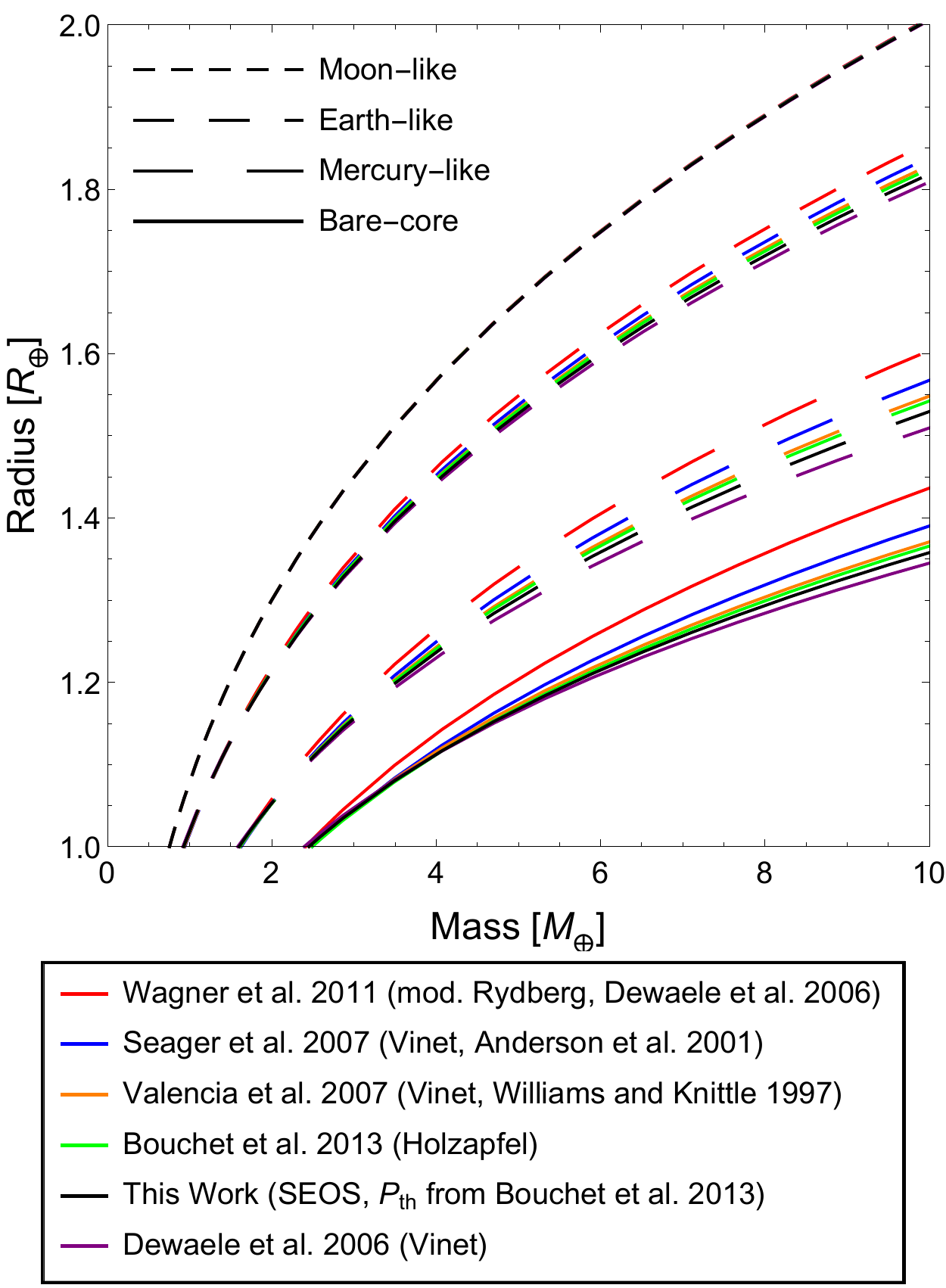}
  \caption[Effect of selected equations of state of Fe on the mass-radius relations of super-Earths]{Effects of selected equations of state of Fe from Table \ref{tab:LiteratureEOSTable} on the mass-radius relations of four classes of super-Earths with fixed core radius fractions. }
  \label{fig:MassRadiusEOS}
\end{figure}

To assess the effect of the choice of an equation of state of iron on the $M$$-$$R$ relations of rocky super-Earths, we have calculated the masses of Moon-like, Earth-like, Mercury-like and bare-core super-Earths with radius between 1$-$2~$\mathrm{R_{\Earth}}$ by solving the interior structure equations (Table \ref{tab:MassEOSTable} and Fig. \ref{fig:MassRadiusEOS}). For the purpose of this argument, the core and mantle compositions are approximated to be pure Fe and $\mathrm{MgSiO_{3}}$. We use $T_{\mathrm{S}}=1650$~K and ignore a temperature jump due to the thermal boundary layer at the CMB. Temperature profiles within the core and mantle are considered to be isentropic. We perform these calculations using SEOS and other equations of state of iron as discussed above.

Central density, temperature and pressure increase with the core radius fraction and with planetary radius (Table \ref{tab:MassEOSTable}). The core mass fractions of Moon-like, Earth-like and Mercury-like super-Earths are in the range 2$-$3\%, 29$-$36\% and 76$-$83\%, respectively. Thermal contributions to total pressure decrease with increasing central pressure. Similarly, the anharmonic-electronic contribution to thermal pressure decreases, but remains at a 10\% level. It decreases the density by 1.6\% at zero pressure and by 0.1\% at 10~TPa. The resulting effects on $M$$-$$R$ relations are not discernible.

Differences in computed planetary masses for different equations of state of iron are negligible for the Moon-like scenarios and $R$=1.25~$\mathrm{R_{\Earth}}$, but they generally increase with increasing central pressure (Table \ref{tab:MassEOSTable}). For the same planetary radius and larger core size, the mass predicted with some equations \citep{Dorn2015,Wagner2011} reach differences by as much as -12\% for a bare core planet. When increasing the planetary radius from 1.25 to 1.75~$\mathrm{R_{\Earth}}$, differences are already noticeable for the smallest core radius fraction, and generally increase by a factor of 2-3 with core size; only for the equation of state by \citet{Dewaele2006} differences increase more strongly. All planetary models based on the equation of state of \citet{Dewaele2006} and some of the 1.25~$\mathrm{R_{\Earth}}$ Earth- and Mercury-type models (based on the equations from \citet{Bouchet2013}, \citet{Fei2016} and \citet{Seager2007}) yield a larger planetary mass due to larger densities over the relevant $P$$-$$T$ range. This difference either stems from the isothermal equation of state (Fig. \ref{fig:EOSIronComparison}) or a smaller thermal pressure contribution than in SEOS. A comparison of differences in $M$$-$$R$ relation over a wider size range (Fig. \ref{fig:MassRadiusEOS}) confirms this trend.

\begin{table*}[!ht]
\caption{\label{tab:BetaEOSTable} Mass-radius relation scaling exponent $\beta$, calculated using different equations of state of Fe. Models assume an isentropic pure~Fe core, an isentropic $\mathrm{MgSiO_{3}}$ mantle, $T_{\mathrm{S}}=1650$~K and a zero CMB $T$-jump. Our calculations for $\beta$ values, using the equations of state from the studies listed below, are performed by fixing the core radius fraction. In the mass-radius relation, $R/R_{\Earth} = a (M/M_{\Earth})^{\beta}$, $a$ is 0.82, 0.89, 1.02 and 1.08 for bare-core, Mercury-like, Earth-like and Moon-like super-Earths, respectively.}
\scriptsize
\begin{center}
  \begin{tabular}{l|SSSS} \hline \hline \rule[0mm]{0mm}{0mm} 
{Equation of state }  & {Bare-core} & {Mercury-like} & {Earth-like} & {Moon-like} \\
$r_{\mathrm{CMB}}$    &           1 &            0.8 &          0.5 & 0.2         \\[2mm]																
\hline 
  SEOS, $P_{\mathrm{th}}$ from \citet{Bouchet2013}      & 0.218 & 0.232 & 0.252 & 0.270 \\
  \citet{Dewaele2006}, Vinet                            & 0.207 & 0.224 & 0.248 & 0.269 \\
  \citet{Bouchet2013}, Holzapfel                        & 0.224 & 0.238 & 0.253 & 0.270 \\  
  \citet{Valencia2007a}, Vinet\textsuperscript{a}       & 0.225 & 0.240 & 0.255 & 0.270 \\
  \citet{Seager2007}, Vinet\textsuperscript{b}          & 0.237 & 0.248 & 0.258 & 0.270 \\
  \citet{Fei2016}, BM3                                  & 0.242 & 0.248 & 0.257 & 0.270 \\
  \citet{Dorn2015}, BM3\textsuperscript{c}              & 0.251 & 0.255 & 0.260 & 0.270 \\
  \citet{Wagner2011}, mod. Rydberg\textsuperscript{d}   & 0.255 & 0.259 & 0.262 & 0.270 \\[2mm]
\hline
  \end{tabular}
  \end{center}\caption*{Abbreviations: $P_{th}$: Thermal Pressure. BM3: third-order Birch-Murnaghan equation. \\ Data source: \textsuperscript{a} \citet{Williams1997}, \textsuperscript{b} \citet{Anderson2001}, \textsuperscript{c} \citet{Belonoshko2010}, \textsuperscript{d} \citet{Dewaele2006}. }
\end{table*}

Looking at the effect of different equation of state formulations for iron on the mass of super-Earths with an Earth-like core ratio, \citet{Wagner2011} found a difference of 2\% in radius for 10~$\mathrm{M_{\Earth}}$, using modified Rydberg and Keane equations, based on the same experimental data of iron from \citet{Dewaele2006}. Interestingly, we find that the use of the Vinet equation of \citet{Dewaele2006}, still based on the same data, leads to a smaller radius by 3\% compared to the modified Rydberg equation from \citet{Wagner2011} for a 10~$\mathrm{M_{\Earth}}$ Earth-like model. As explained above, this discrepancy stems from their choice of a specific value for the derivative of the bulk modulus at infinite pressure.

The differences in radii calculated using the two equations of state of iron with maximum density difference are 1\% and 2\% ($M$=1~$\mathrm{M_{\Earth}}$), 2\% and 4\% ($M$=5~$\mathrm{M_{\Earth}}$), and 3\% and 6\% ($M$=10~$\mathrm{M_{\Earth}}$) for Earth-like and Mercury-like super-Earths, respectively. Such differences result in significant variations for the exponent $\beta$ in the mass-radius scaling law (equation \ref{eq:ScalingLaw}), and values for $\beta$ can vary by as much as 10\% (Table \ref{tab:BetaEOSTable}).  These differences are a consequence of extrapolations of the equations meant to be used up to the pressures of $\sim$0.3~TPa. Since the Holzapfel equation from \citet{Bouchet2013} needs to be extrapolated starting at 1.5~TPa, its $M$$-$$R$ curves diverge significantly from those based on SEOS for super-Earths with larger cores or larger planetary radii only (Fig.~\ref{fig:MassRadiusEOS}).

\section{Effects of Temperature and Composition on M-R Relations}\label{paper1:effectOthers}

\subsection{Temperature}\label{paper1:effectTemp}

Since the temperature profile in super-Earths is not well constrained, it is important to assess the effect of uncertainties in temperature on the structure of super-Earths. Previous studies indicate a relatively minor effect \citep[e.g.,][]{Sotin2007,Grasset2009}. Here we re-evaluate this effect for some of the temperature assumptions described in Section \ref{paper1:method}.

For Moon-like models, a change in the temperature profile in the core from isentropic to isothermal reduces the central temperature by 200$-$600~K and has a negligible impact on mass. For Earth-like and Mercury-like models, the differences between the isentrope and isotherm in the center are 1000$-$4500~K and 2500$-$10000~K and masses by up to 1\% and 2\%, respectively. 

\begin{figure}[!ht]
  \centering
  \medskip
  \includegraphics[width=.45\textwidth]{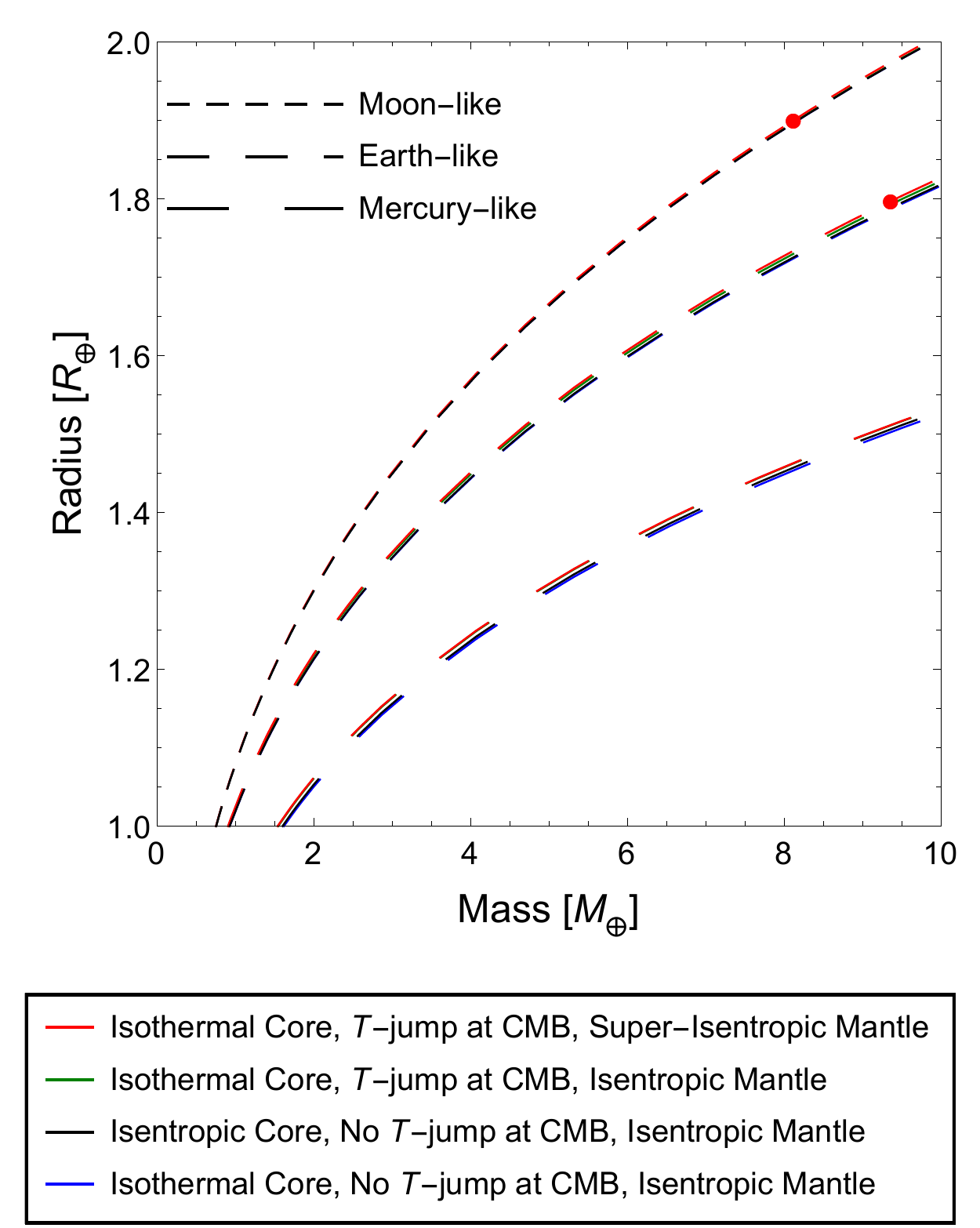}
  \caption[Effect of temperature on the mass-radius relation of super-Earths]{Effects of temperature on mass-radius relations of super-Earths with fixed core radius fractions. Red dots on red lines represent the mass and radius where the CMB temperature is at the solvus. For Mercury-like super-Earths, the cut-off occurs for planets with $M$~{\textgreater}~10~$\mathrm{M_{\Earth}}$. }
  \label{fig:MassRadiusTempComp}
\end{figure}

An extreme CMB temperature jump of around 2500$-$7000~K (see Section \ref{paper1:method}) reduces the density of the core, it has a negligible impact on Moon-like models, but decreases the mass of Earth-like and Mercury-like super-Earths between 1$-$2\% and 2$-$5\%, respectively. This temperature jump effect decreases with planetary radius because thermal contribution to pressure decreases. Conversely, for a given planetary mass, the planetary radius increases by up to 1\% and 1.5\% for Earth-like and Mercury-like models (Fig.~\ref{fig:MassRadiusTempComp}). These results are comparable to those of \citet{Grasset2009} who found a difference of up to 2\% in radius when varying the planetary temperature profiles by 4000~K. To summarize, the effect of the core thermal profile and the CMB temperature jump on $M$$-$$R$ relations increases with core size.

In the lower mantle, a super-adiabatic thermal gradient, starting at the transition to Mg-ppv, leads to a decrease in the mass of super-Earths by up to 3\%. This effect is most prominent for moderate size cores (Earth-like) since smaller or larger cores lead to a smaller super-adiabatic profile in the mantle because the pressure either increases more slowly with depth or the mantle is thinner. In the super-adiabatic case, if the temperature in the lower mantle is higher than the solvus temperature of MgO, parts of the lower mantle might be soluble in the iron core \citep{Wahl2015}. Our calculations for a pure Fe core show that the miscibility of rock and iron can occur when $R_{\mathrm{Moon-like}}$~{\textgreater}~1.9~$\mathrm{R_{\Earth}}$, $R_{\mathrm{Earth-like}}$~{\textgreater}~1.8~$\mathrm{R_{\Earth}}$ and $R_{\mathrm{Mercury-like}}$~{\textgreater}~1.75~$\mathrm{R_{\Earth}}$ (see Fig.~\ref{fig:MassRadiusTempComp}). For such planets, the core size is defined by the solvus. If the core contains light elements, the pressure at the CMB for a given planet radius is lower than in a planet with a pure~Fe core, therefore a dissolution of the lower mantle in the core occurs for planets with larger radii than for the pure~Fe core case.

Overall, the effect of temperature on mass-radius relations is smaller than the effect of extrapolations of the equations of state of iron as seen in Figs.~\ref{fig:MassRadiusEOS} and \ref{fig:MassRadiusTempComp}. Mass-radius relations due to changes in temperature profile differ only by $\Delta \beta$ {\textless} 0.003 for all cases considered here. This confirms the results obtained by e.g., \citet{Sotin2007} and \citet{Grasset2009} and extends their validity range. Temperature, nevertheless, is important for many dynamic aspects of a planet, such as convection, volcanism as well as potential habitability. 

\begin{table*}[!ht]
\caption{\label{tab:MassCoreCompTable} Differences in planetary mass of super-Earths with $R$=1.25~$\mathrm{R_{\Earth}}$ and $R$=1.75~$\mathrm{R_{\Earth}}$ between different core compositions and core sizes. All differences in mass are calculated with respect to the model with a pure~Fe core. Models assume an isentropic core, an isentropic $\mathrm{MgSiO_{3}}$ mantle, $T_{\mathrm{S}}=1650$~K and a zero CMB $T$-jump. }
\scriptsize
\begin{center}
  \begin{tabular}{ll|cccc} \hline \hline \rule[0mm]{0mm}{0mm} 
{Core comp.} & Parameter  & {Bare-core} & {Mercury-like} & {Earth-like} & {Moon-like} \\
& $r_{\mathrm{CMB}}$            & 1	& 0.8 & 0.5 & 0.2  \\[2mm]	
\hline
\multicolumn{2}{c}{$R$ = 1.25 $\mathrm{R_{\Earth}}$} & & & & \\
0.9$\rho$ Fe     & $\Delta M$ & $-$21\% & $-$16\%  & $-$6\%   & $-$0.5\% \\
0.8$\rho$ Fe     & $\Delta M$ & $-$38\% & $-$28\%  & $-$11\%  & $-$0.6\%   \\
FeS               & $\Delta M$ & $-$40\% & $-$29\%  & $-$11\%  & $-$0.6\%   \\
FeSi              & $\Delta M$ & $-$40\% & $-$30\%  & $-$11\%  & $-$0.6\%   \\
Fe$_{0.95}$O      & $\Delta M$ & $-$43\% & $-$31\%  & $-$12\%  & $-$1.2\%   \\
Fe$_{3}$C         & $\Delta M$ & $-$11\% & $-$9\%   & $-$4\%   & $-$0.3\% \\
\multicolumn{2}{c}{$R$ = 1.75 $\mathrm{R_{\Earth}}$}   & & & & \\
0.9$\rho$ Fe     & $\Delta M$ & $-$29\%  & $-$22\%   & $-$8\%    & $-$0.5\%   \\
0.8$\rho$ Fe     & $\Delta M$ & $-$50\%  & $-$39\%   & $-$15\%   & $-$1.0\%   \\
FeS               & $\Delta M$ & $-$52\%  & $-$39\%   & $-$15\%   & $-$1.0\%   \\
FeSi              & $\Delta M$ & $-$51\%  & $-$39\%   & $-$15\%   & $-$1.0\%   \\
Fe$_{0.95}$O      & $\Delta M$ & $-$49\%  & $-$37\%   & $-$14\%   & $-$1.0\%   \\
Fe$_{3}$C         & $\Delta M$ & $-$8\%   & $-$5\%    & $-$2\%    & $-$0.2\%   \\[2mm]
\hline
  \end{tabular}
  \end{center}
\end{table*}

\subsection{Core Composition}\label{paper1:effectCoreComp}

\begin{figure}[!ht]
  \centering
  \medskip
  \includegraphics[width=.48\textwidth]{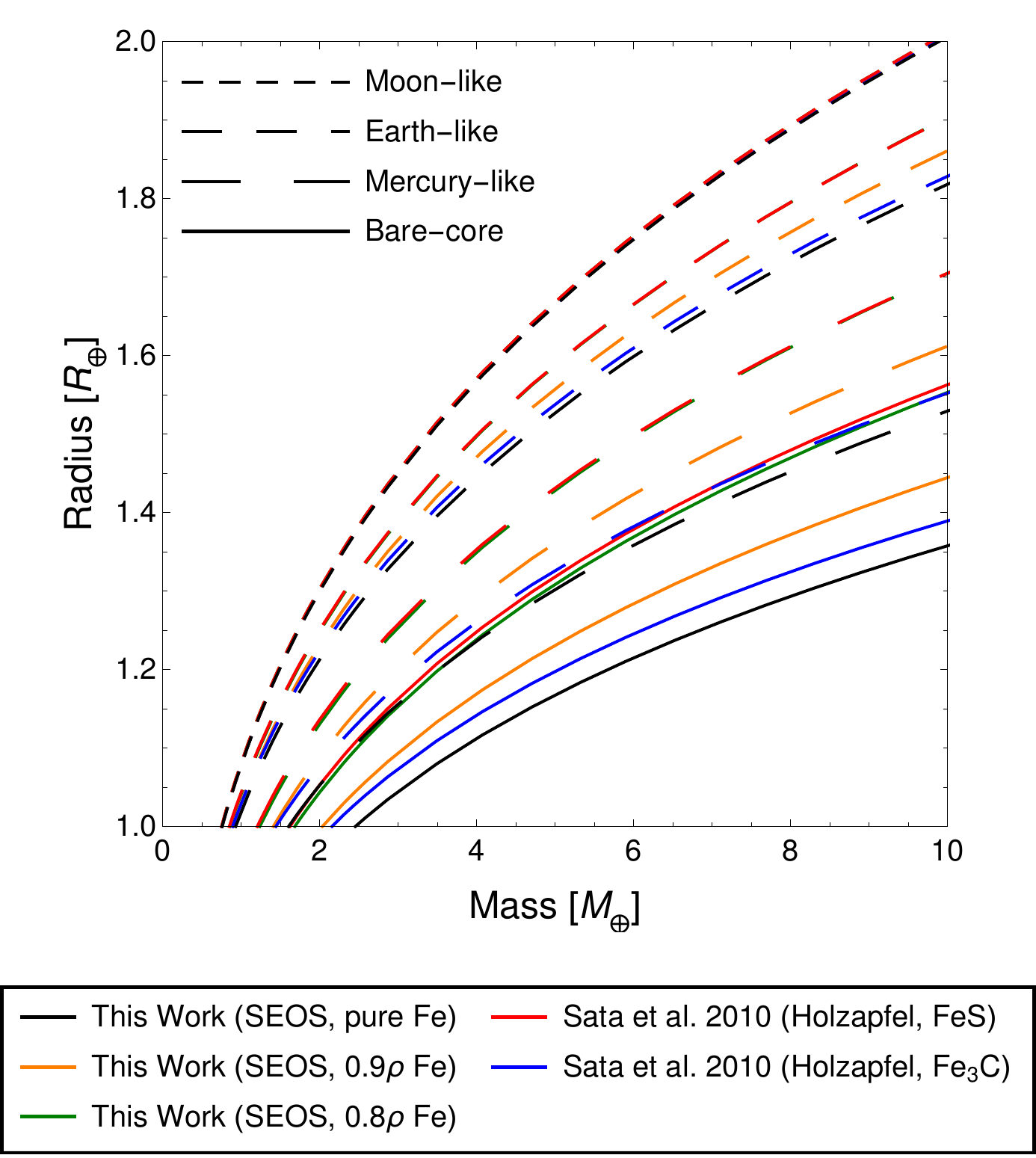}
  \caption[ Effect of different core compositions on the mass-radius relations of super-Earths]{Effect of different core compositions on mass-radius relations of super-Earths with fixed core radius fractions. The Moon-like, Earth-like and Mercury-like models for our 0.8$\rho$ Fe case and the FeS \citep{Sata2010} case overlap with each other. \citet{Sata2010} measured the equation of state data of FeS and Fe$_{3}$C up to $\sim$200~GPa.} 
  \label{fig:MassRadiusCoreComp}
\end{figure}

The composition of the core is one of the prime factors determining the mass of a super-Earth with a given radius \citep[e.g.,][]{Unterborn2016}. \citet{Valencia2007a} showed that the presence of a relatively small amount of light elements in the core, smaller than the amount in the Earth's core, can change the radius of large super-Earths with given mass by a non-negligible amount. Here we quantify this effect for prescribed density deficits with respect to pure iron and compare results obtained for different core compositions. The effect on mass-radius relations is negligible for Moon-like super-Earths, but significant for planets with larger cores (Fig.~\ref{fig:MassRadiusCoreComp}). 

Models with Fe$_{3}$C cores are lighter than those with pure~Fe cores resulting in smaller masses (Table \ref{tab:MassCoreCompTable}). Super-Earths with 0.9$\rho$~Fe cores are denser than FeS but lighter than Fe$_{3}$C. Models with FeS, FeSi or Fe$_{0.95}$O cores have similar masses. Mass-radius curves with 0.8$\rho$~Fe as core composition are almost overlapping with those of FeS, FeSi and Fe$_{0.95}$O as core compositions, implying that the addition of 50 mol\% (36 wt\%) S, 50 mol\% (33.5 wt\%) Si or 51 mol\% (23 wt\%) O in the Fe-core is virtually equivalent to a 20\% density deficit with respect to pure~Fe. We do not show the overlapping $M$$-$$R$ curves of the FeSi and Fe$_{0.95}$O cases in Fig.~\ref{fig:MassRadiusCoreComp}. It is interesting to note that the mass-radius curve of Mercury-like super-Earth with a pure~Fe core and a $\mathrm{MgSiO_{3}}$ mantle overlaps with that of FeS bare-core model, implying it is impossible to distinguish between the two with the knowledge of mass and radius only.
	
\subsection{Mantle Composition}\label{paper1:effectMantleComp}

\begin{table*}[!ht]
\caption{\label{tab:MassMantleCompTable} Differences in planetary mass of super-Earths with $R$=1.25~$\mathrm{R_{\Earth}}$ and $R$=1.75~$\mathrm{R_{\Earth}}$ between different mantle compositions and core sizes. All differences in mass are calculated with respect to the model with a $\mathrm{MgSiO_{3}}$ mantle. Models assume and isentropic pure~Fe core, an isentropic mantle, $T_{\mathrm{S}}=1650$~K and a zero CMB $T$-jump. }
\scriptsize
\begin{center}
  \begin{tabular}{ll|cccc} \hline \hline \rule[0mm]{0mm}{0mm} 
{Mantle comp.} & Parameter & {Mercury-like} & {Earth-like} & {Moon-like} & {Coreless}  \\
& $r_{\mathrm{CMB}}$            & 0.8	& 0.5 & 0.2 & 0  \\[2mm]	
\hline 
\multicolumn{2}{c}{$R$ = 1.25 $\mathrm{R_{\Earth}}$} & & & & \\
$\mathrm{Mg_{2}SiO_{4}}$ & $\Delta M$ & $-$0.8\%  & $-$1.4\%  & $-$2.4\%  & $-$2.4\%  \\
$\mathrm{SiO_{2}}$       & $\Delta M$ & $+$1.0\%  & $+$2.3\%  & $+$2.9\%  & $+$3.0\%  \\
$\mathrm{FeSiO_{3}}$     & $\Delta M$ & $+$11\%   & $+$32\%   & $+$44\%   & $+$58\%   \\
$\mathrm{Fe_{2}SiO_{4}}$ & $\Delta M$ & $+$15\%   & $+$44\%   & $+$44\%   & $+$59\%   \\
\multicolumn{2}{c}{$R$ = 1.75 $\mathrm{R_{\Earth}}$} & & & & \\
$\mathrm{Mg_{2}SiO_{4}}$ & $\Delta M$ & $-$0.5\%  & $-$0.7\%  & $-$1.2\%  & $-$1.2\%  \\
$\mathrm{SiO_{2}}$       & $\Delta M$ & $+$3.0\%  & $+$1.2\%  & $+$1.5\%  & $+$1.7\%  \\
$\mathrm{FeSiO_{3}}$     & $\Delta M$ & $+$15\%   & $+$48\%   & $+$63\%   & $+$64\%   \\
$\mathrm{Fe_{2}SiO_{4}}$ & $\Delta M$ & $+$20\%   & $+$63\%   & $+$82\%   & $+$83\%   \\[2mm]
\hline
  \end{tabular}
  \end{center}
\end{table*}

\begin{figure}[!ht]
  \centering
  \medskip
  \includegraphics[width=.45\textwidth]{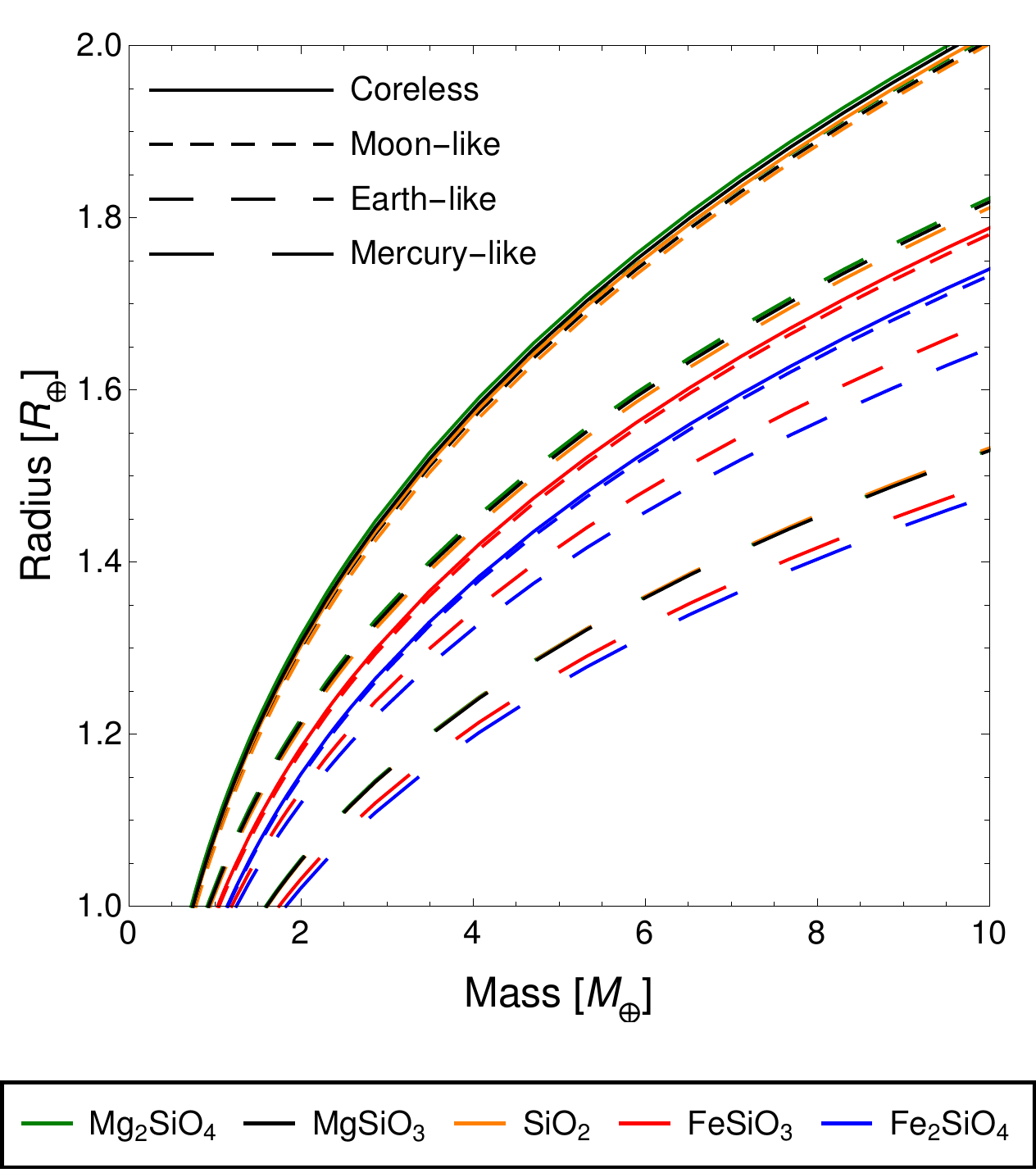}
  \caption[Effect of mantle composition on the mass-radius relations of super-Earths]{Effect of different mantle compositions on mass-radius relations of super-Earths with fixed core radius fractions. Lines corresponding to $\mathrm{MgSiO_{3}}$, $\mathrm{Mg_{2}SiO_{4}}$ and $\mathrm{SiO_{2}}$ almost overlap with each other, with $\mathrm{SiO_{2}}$ being the most dense and $\mathrm{Mg_{2}SiO_{4}}$ being the least dense among these three compositions.}
  \label{fig:MassRadiusMantleComp}
\end{figure}

Although mantle composition is mostly considered to be of lesser importance for the mass-radius relation of super-Earth exoplanets than core composition \citep[e.g.,][]{Unterborn2016}, only a limited range of mantle compositions has been considered previously. A reasonable range of mantle compositions based on stellar elemental ratios has been considered in \citet{Dorn2015}, but these authors do not investigate their effect on mass-radius relations. As explained in Section \ref{paper1:method}, we here evaluate the effect of a wide range of mantle compositions and use the thermodynamic code of \citet{Chust2017} for the calculations of the various mineral phase transitions with increasing pressure. The effect of mantle composition on mass-radius relations increases with decreasing core size and increasing planetary radius (Fig.~\ref{fig:MassRadiusMantleComp}). Moon-like and coreless models have almost indistinguishable mass-radius relations because of the small cores of Moon-like models. 

Models with mantle compositions of $\mathrm{MgSiO_{3}}$, $\mathrm{Mg_{2}SiO_{4}}$ and $\mathrm{SiO_{2}}$ differ by less than 2\% (Table \ref{tab:MassMantleCompTable}) because of the similarity in densities of MgO-SiO$_{2}$-based minerals. This finding extends that of \citet{Grasset2009} who showed that the effect of different Mg/Si ratios on planetary radius is {\textless} 2\%, but did not consider high-pressure mineral phase transitions of perovskite. Our calculations show that post-perovskite is the most abundant mineral in all Mg-rich super-Earths larger than the Earth. The dissociated assemblages of Mg-ppv \citep{Umemoto2011}, $\mathrm{MgSi_{2}O_{5}}$ + MgO and $\mathrm{Fe_{2}P}$-type $\mathrm{SiO_{2}}$ + MgO, would be present in planets larger than $R\sim$1.7~$\mathrm{R_{\Earth}}$ and $R\sim$1.9~$\mathrm{R_{\Earth}}$, respectively. The inclusion of the dissociated phases of ppv has a negligible effect on the mass-radius relations of super-Earths.

The presence of FeO~end-member minerals in the mantle increases the mass of the planet by a large amount, depending on the core size and the planetary radius (Table \ref{tab:MassMantleCompTable}). For an Earth-like planet with $R$=1~$\mathrm{R_{\Earth}}$, switching the mantle composition from $\mathrm{MgSiO_{3}}$ to $\mathrm{FeSiO_{3}}$ leads to a higher mass by 26\%, comparable to the result of \citet{Unterborn2016}. Even though FeO~end-members do not represent realistic mantles, they give an upper-limit and illustrate the large effect of the Fe-content of the mantle on mass-radius relations.

\subsection{M-R Degeneracy and Observed Super-Earths}\label{paper1:massRadiusObserved}

\begin{figure*}[!ht]
  \centering
  \medskip
  \includegraphics[width=0.95\textwidth]{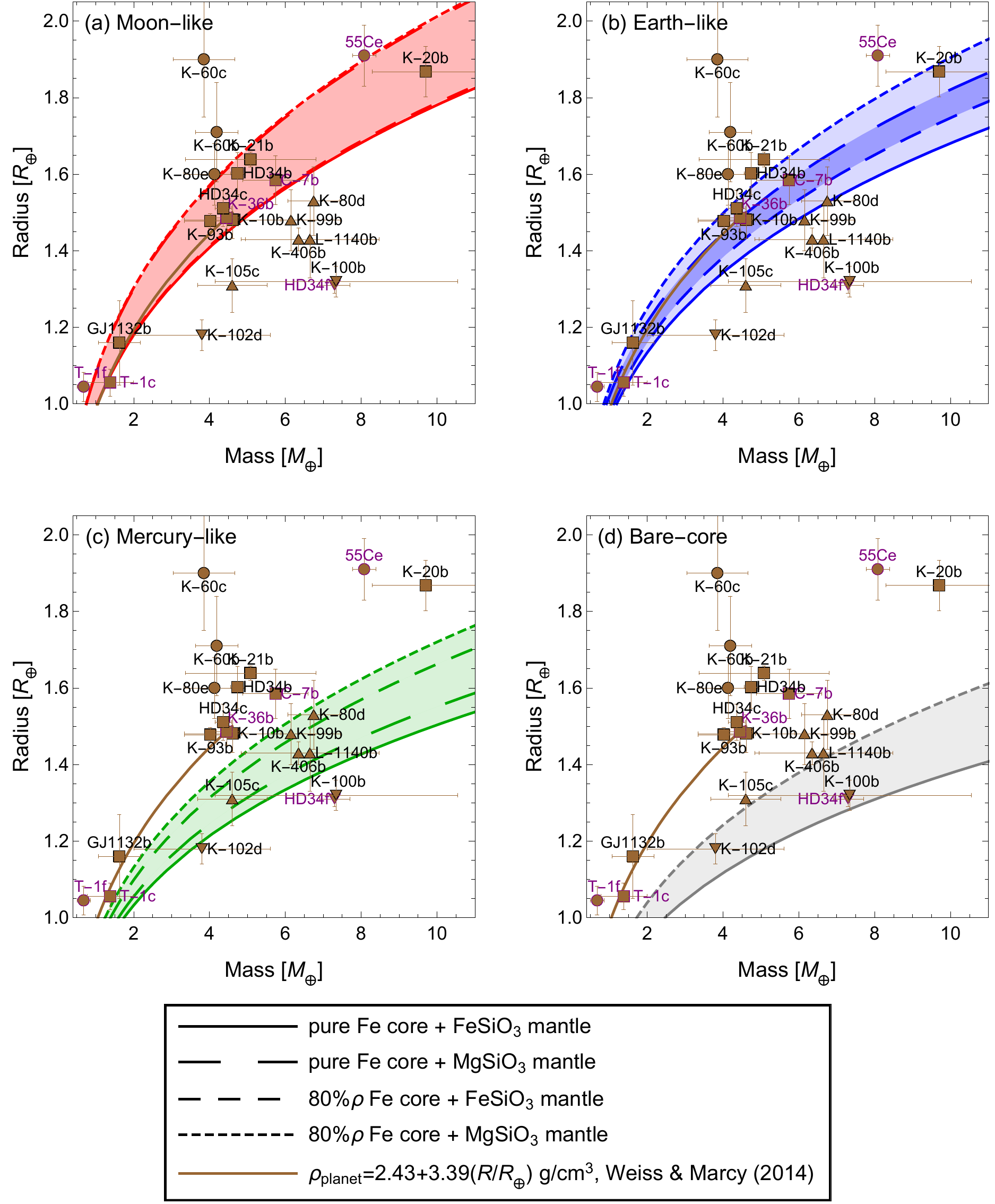}
  \caption[Observed super-Earths in the context of theoretical mass-radius relations]{$M$-$R$ relations of bare-core, Mercury-, Earth- and Moon-like super-Earths with fixed core radius fractions. Core compositions are 0.8$\rho$ Fe or pure Fe and mantle compositions of $\mathrm{MgSiO_{3}}$ or $\mathrm{FeSiO_{3}}$, although $\mathrm{FeSiO_{3}}$ is much less likely. Observed exoplanets shown as squares have core radius fraction similar to the Earth or smaller, triangles have core radius fraction much larger than the Earth, inverted triangles have almost naked cores and circles are likely composed of significant amounts of volatile material or gas envelopes. Planets highlighted in violet are discussed in the text. Notation: T: TRAPPIST, K: Kepler, C: CoRoT, 55Ce: 55~Cnc~e, HD34: HD~219134, L: LHS.}
  \label{fig:MassRadiusObs}
\end{figure*}

Uncertainties in mantle and core composition can introduce large errors in the determination of the core size of a rocky super-Earth without a gaseous envelope. In order to illustrate and quantify the combined effect of mantle and core compositions, we plot in Fig.~\ref{fig:MassRadiusObs} the mass-radius relations of bare-core, Mercury-like, Earth-like and Moon-like super-Earths with two extreme core compositions (pure Fe and 0.8$\rho$ Fe) and two extreme mantle compositions ($\mathrm{MgSiO_{3}}$ and $\mathrm{FeSiO_{3}}$). Since the effect of temperature is small, we restrict ourselves to one temperature profile. We assume an isentropic core with no temperature jump at the CMB. We also plot observed $M$$-$$R$ curves for planets with $R < 1.5~\mathrm{R_{\Earth}}$ from \citet{Weiss2014} and observed super-Earths\footnote{http://exoplanetarchive.ipac.caltech.edu}, having uncertainties on radii and masses less than 10\% and 50\%, respectively. 

In each of the four classes of super-Earths considered, the heaviest planets have a pure~Fe core with a $\mathrm{FeSiO_{3}}$ mantle and the lightest planets have a 0.8$\rho$ ~Fe~core with a $\mathrm{MgSiO_{3}}$ mantle. The colored bands in the $M$$-$$R$ plane in Fig.~\ref{fig:MassRadiusObs} represent the spread due to variations in mantle and core compositions. For Mercury-like super-Earths, the planets with a pure~Fe core and a $\mathrm{MgSiO_{3}}$ mantle are heavier than the planets with a 0.8$\rho$ ~Fe~core and a $\mathrm{FeSiO_{3}}$ mantle. However, the opposite is true for Earth-like and Moon-like planets which have smaller core size (Fig.~\ref{fig:MassRadiusObs}a$-$c). 

As an example of the degeneracy in estimating the core size due to uncertainties in mantle and core compositions we consider the case of CoRoT-7b. Its position in Fig.~\ref{fig:MassRadiusObs}(b) on the Earth-like mass-radius curve with a pure Fe core and a $\mathrm{MgSiO_{3}}$ mantle suggests that the core radius fraction of CoRoT-7b is about $\sim$0.5, for a pure~Fe core and a $\mathrm{MgSiO_{3}}$ mantle. A Mercury-like core size can be ruled out based on Fig.~\ref{fig:MassRadiusObs}(c). On the other hand, given the position of CoRoT-7b in Fig.~\ref{fig:MassRadiusObs}(a), it can also be Moon-like (r$_{\mathrm{CMB}}$ $\sim$0.2) with a pure~Fe core and a mantle in composition close to $\mathrm{FeSiO_{3}}$. Uncertainties on the mass and radius of CoRoT-7b (15\% and 4\%, respectively) further contribute to the difficulty of constraining the core size. Additional information about composition, perhaps from the stellar composition as suggested by \citet{Dorn2015}, is needed to lift the degeneracy. 

For super-Earths with more extreme locations on the $M$$-$$R$ diagram, the degeneracy can be weaker. For example, HD~219134f, which is more massive than the Mercury-like or 0.8$\rho$~Fe bare-core super-Earths (see Fig.~\ref{fig:MassRadiusObs}(c) and (d)), is likely the stripped-off core of a giant planet as suggested for some other planets by \citet{Mocquet2014}. Super-Earths that lie above the mass-radius bands, even with the lightest pure $\mathrm{MgSiO_{3}}$ composition, e.g., TRAPPIST-1f, have a lower density than a planet made purely of silicates (Fig.~\ref{fig:MassRadiusObs}(a)) and likely contain significant amounts of lighter material such as ice or water in addition to rock. Another super-Earth in the same planetary system, TRAPPIST-1c, is situated in the denser part of the band of Earth-like models, and likely has a core radius fraction between that of the Earth and Mercury, assuming it has an Earth-like composition in the core and the mantle. The uncertainties on its mass, however, are too large to confidently make inferences on its interior.

55~Cancri~e and the planets in the Kepler-60 system are also lighter than the lightest rocky planets considered here (Fig.~\ref{fig:MassRadiusObs}). This observation implies that such low-mass planets have thick atmospheres, which increase their radii by significant amounts as suggested by, e.g., \citet{Fortney2007}. However, the presence of large amounts of lighter material such as water, ices or carbon compounds, not considered in this study, can also explain their low mean density, increasing the degeneracy \citep[e.g.,][]{Seager2007,Valencia2007a,Grasset2009}.

\section{Uncertainties on the Interior Structure of Kepler-36b}\label{paper1:interiorKepler36b}

\begin{figure*}[!ht]
  \centering
  \medskip
  \includegraphics[width=0.95\textwidth]{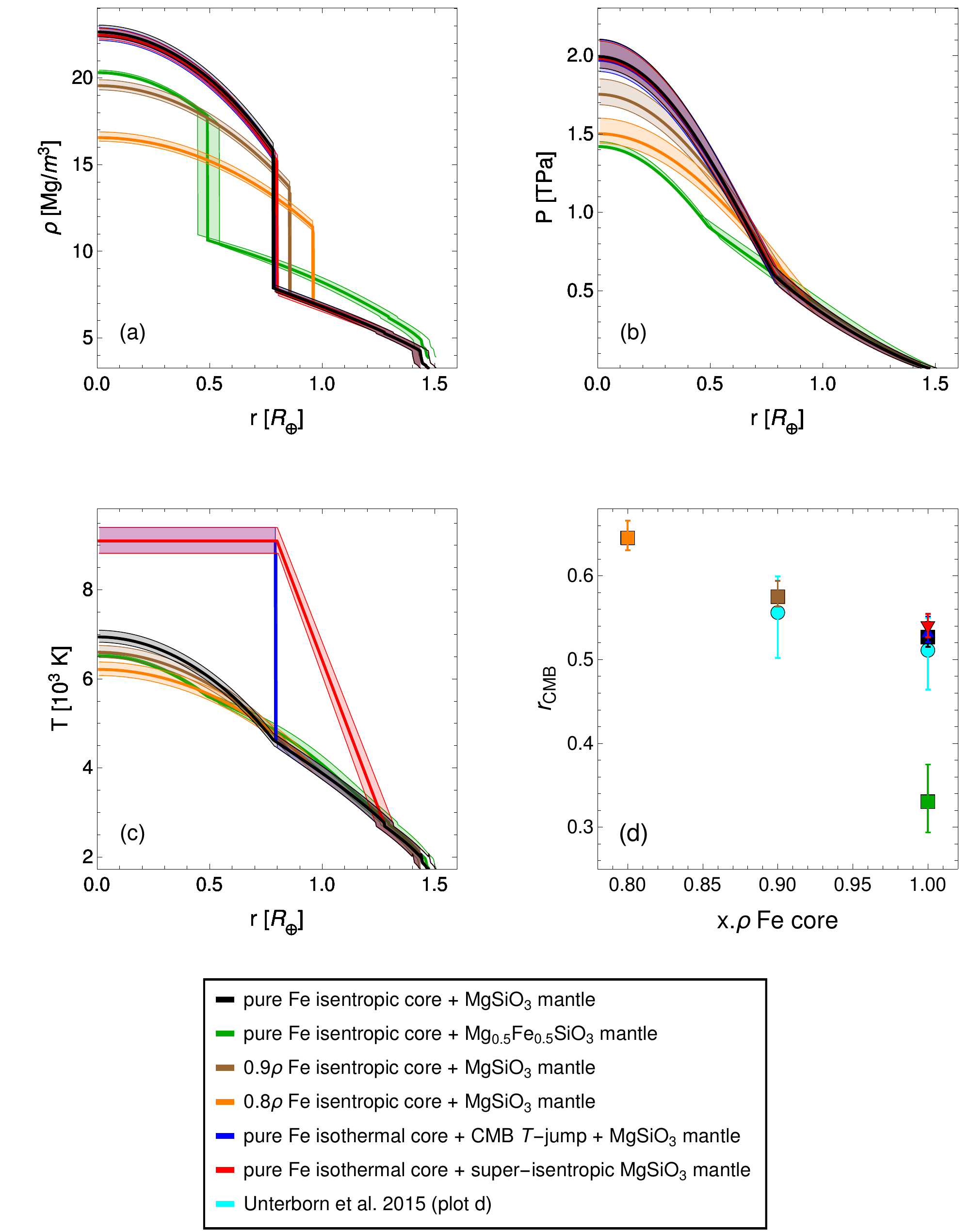}
  \caption[The interior structure of Kepler-36b]{Density, pressure and temperature distribution, and the core radius fraction of Kepler-36b for selected modeling scenarios. Shaded areas represent uncertainties arising due to uncertainties in mass and radius of Kepler-36b. }
  \label{fig:Kepler36bPlot}
\end{figure*}

In this section we compute in detail the interior density, pressure and temperature profiles and the core radius fraction ($r_{\mathrm{CMB}}$) for one example, Kepler-36b, a rocky super-Earth with the smallest uncertainty on its observed mass and radius, $M=4.45^{+0.33}_{-0.27}$ $\mathrm{M_{\Earth}}$ and $R=1.486 \pm 0.035$ $\mathrm{R_{\Earth}}$ \citep{Carter2012}. Assuming a pure Fe core and an $\mathrm{MgSiO_{3}}$ mantle and using the SEOS for Fe, $r_{\mathrm{CMB}}$ of Kepler-36b is $0.527^{+0.018}_{-0.012}$, implying it is an Earth-like planet by our definition (Fig.~\ref{fig:Kepler36bPlot}(d)). Use of equations of state of Fe other than SEOS changes $r_{\mathrm{CMB}}$ by $-1$\% to 4\% (see Fig.~\ref{fig:Kepler36bPlotEOS}). The associated relative differences in density and pressure profiles reach 15\%. Density increases monotonically from the surface to the center due to adiabatic self-compression, with discontinuous jumps at the mantle phase transitions, dominated by the perovskite and ppv forming transitions. Phases beyond ppv \citep{Umemoto2011} do not occur because the pressures in the mantle of Kepler-36b are not high enough for the dissociation. 

A change of planetary temperature profiles has a small effect on the interior properties of Kepler-36b (Fig.~\ref{fig:Kepler36bPlot}(d)). Assuming an isothermal core with a CMB temperature jump of 9000~K, determined by the melting temperature of MgSiO$_{3}$ at 0.6~TPa \citep{Belonoshko2005}, increases $r_\mathrm{CMB}$ by 1\% (Fig.~\ref{fig:Kepler36bPlot}(d)). A super-adiabatic thermal gradient in the lower mantle further increases $r_\mathrm{CMB}$ by 2\%. Core densities and pressures decrease by up to 2\% because of a hotter and lighter interior. Since the super-adiabatic temperatures are below the solvus temperature of MgO \citep{Wahl2015}, we do not expect high-temperature miscibility of iron and rock in the interior of Kepler-36b.

The presence of light elements in the core has a significant effect on the interior properties of Kepler-36b. Considering core compositions of 0.9$\rho$~Fe and 0.8$\rho$~Fe (equivalent to FeS), we predict $r_\mathrm{CMB}$ of Kepler-36b to increase by 9\% and 22\% compared to the case of pure~Fe. The central density and pressure decrease by $\sim$13\% ($\sim$25\%) for 0.9$\rho$~Fe (0.8$\rho$~Fe). The combination of a MgSiO$_{3}$ mantle and a 0.8$\rho$~Fe core yields the largest core for Kepler-36b within our model scenarios. 

Using different MgO-SiO$_{2}$-based mantle compositions ($\mathrm{Mg_{2}SiO_{4}}$ and $\mathrm{SiO_{2}}$) show small effects on the core radius (1\% and $-$2\%, respectively). Switching to FeO-based end-members, i.e., $\mathrm{FeSiO_{3}}$ and $\mathrm{Fe_{2}SiO_{4}}$, the interior of Kepler-36b cannot be modeled successfully, because resulting densities are so high that the resulting mass exceeds the measured value for Kepler-36b. For instance, a pure $\mathrm{FeSiO_{3}}$ planet with no core and a radius of 1.486 $\mathrm{R_{\Earth}}$ would have a mass of 4.7 $\mathrm{M_{\Earth}}$ which exceeds the mass of Kepler-36b by 6\%. 

With a mid-member mantle composition of $\mathrm{Mg_{0.5}Fe_{0.5}SiO_{3}}$ we find $r_{\mathrm{CMB}}=0.330^{+0.045}_{-0.036}$, which is smaller than the case of $\mathrm{MgSiO_{3}}$ by 37\%. In the core, the density, pressure and temperature decrease by as much as 25\%, while in the mantle there is a significant increase only in density, but not pressure or temperature (Fig.~\ref{fig:Kepler36bPlot}). Theoretically, the mass of Kepler-36b can be explained without any core, by considering an undifferentiated composition of about $\mathrm{Mg_{0.15}Fe_{0.85}SiO_{3}}$. 

Taking all observational and modeling uncertainties into account, it is possible to give an upper bound to the core radius of Kepler-36b of $r_{\mathrm{CMB}}=0.67$. For the models considered here with both the core and mantle, the central density and the central pressure belong in the range 16.3$-$22.9~$\mathrm{Mg/m^{3}}$ and 1.4$-$2.2~TPa. For our coreless model, the central density and the central pressure are 11.2~$\mathrm{Mg/m^{3}}$ and 1.0~TPa. The differences in the interior properties of Kepler-36b due to modeling uncertainties illustrate that temperature and equation of state of iron play a minor role on planetary interiors, compared to composition. As Fig.~\ref{fig:Kepler36bPlot} clearly demonstrates, modeling uncertainties dominate over observational uncertainties on the interior properties of Kepler-36b. 
 
\section{Summary and Conclusions}\label{paper1:discussion}

Previous studies \citep[e.g.,][]{Seager2007,Grasset2009,Wagner2011} have pointed to the uncertainty of the equations of state of iron at pressures relevant to super-Earth cores (above 0.3~TPa) due to a lack of experimental data. In this paper, we derive a new ab initio equation of state of iron for super-Earths (SEOS) based on Density Functional Theory (DFT) calculations. In addition to the discrete set of ab initio results given in the supplement, we also provide a closed equation of state expression by fitting the DFT results to the Holzapfel equation to 10~TPa. 

Density differences between SEOS and other equations of state of iron from the literature are restricted to $\pm 2$\% up to $\sim$0.5~TPa, but range between $-$20\% and 5\% at 10~TPa. The equation of state from \citet{Bouchet2013} based on the Holzapfel formulation has the smallest density difference with SEOS ({\textless} 3.5\%) up to 10~TPa. These density differences have significant effects on mass-radius relations of rocky super-Earths without gaseous envelopes, in particular for planets with an iron-core large enough to reach pressures where extrapolations of equations of state of iron are required. For models with an Earth-like core radius fraction of 0.5, the use of an inadequate equation of state can result in a difference in the mass of up to 10\% for a 2~$\mathrm{R_{\Earth}}$ super-Earth, and a difference in the radius of up to 3\% for a 10~$\mathrm{M_{\Earth}}$ super-Earth. For models with Mercury-like core radius fraction of 0.8 and bare-core models, these effects are even larger, reaching 20\% in mass and 6\% in radius. For models with Moon-like core radius fraction of 0.2, the core pressures are not high enough to induce significant differences in mass-radius relations. 

Although the effects of temperature on mass-radius relations are smaller than those of the equation of state for iron, assumptions of an extreme CMB temperature jump or a super-adiabatic temperature profile in the lower mantle can change the mass of super-Earths by up to 5\%, depending on the core size. We find that mantle temperatures, when assuming a super-adiabatic profile, are too low to allow for rock and iron to mix for planets smaller than 1.75~$\mathrm{R_{\Earth}}$. Light elements in the core can strongly reduce the density in the core and therefore the mass of an exoplanet. By assuming a FeS core (equivalent to a reduction in density with respect to iron of 20\%) instead of pure~Fe, mass decreases by up to 13\% and 33\% for Earth- and Mercury-like super-Earths, respectively. The impact of the Mg/Si ratio on the mass-radius relations is very small because of the similarity in densities of MgO-SiO$_{2}$-based minerals. However, the presence of FeO in the mantle has a significant effect, as Fe-bearing minerals have higher densities than Mg-bearing minerals. Assuming a mantle composition of pure $\mathrm{FeSiO_{3}}$, although much less likely than $\mathrm{MgSiO_{3}}$, for a fixed mass, decreases the radius by up to 8\% and 10\% for Earth- and Moon-like super-Earths, respectively.

To quantify the effects of modeling uncertainties on the interior structure, we use Kepler-36b, a super-Earth with well-constrained mass and radius, as a test case. We demonstrate that the uncertainties due to the equation of state of iron (5\%), temperature (2\%), core and mantle composition ({\textgreater} 20\%) on the core radius of Kepler-36b dominate over the observational uncertainties on its mass and radius, which are 7\% and 2\%, respectively. Similar modeling uncertainties are observed for other interior properties such as central density and pressure.

We show that the use of an appropriate equation of state of iron (\ref{paper1:appendixExtrapolation}) reduces the degeneracy in interpretations of mass and radius, but that uncertainties in composition, and to a minor extent temperature, lead to a spread of mass-radius curves into bands for Moon-, Earth- and Mercury-like exoplanets that can overlap. This significantly limits the ability to accurately infer the interior structure of rocky super-Earths. For example, the mass and radius of CoRoT-7b can equally be satisfied by a pure~Fe core and a $\mathrm{MgSiO_{3}}$ mantle for an Earth-like core or a pure~Fe core and a $\mathrm{FeSiO_{3}}$ mantle for a Moon-like core. This non-uniqueness adds to the large degeneracy that may result from not knowing if the planet has a substantial gaseous envelope. Since masses and radii of rocky super-Earths are expected to be determined with an accuracy as high as 3\% in the planet radius and 10\% in the planet mass \citep{Hatzes2016} with upcoming missions such as TESS, CHEOPS, JWST and PLATO, modeling uncertainties will then dominate over observational uncertainties. Knowledge of the stellar composition, for example Fe/Si and Mg/Si ratios, would play an important role in mitigating the $M$$-$$R$ degeneracy \citep{Dorn2015,Santos2017}.

\section*{Acknowledgments}
We thank Diana Valencia and an anonymous reviewer for their insightful comments in improving this manuscript. This research has been supported by the Planetary and Exoplanetary Science Network (PEPSci), funded by the Netherlands Organization for Scientific Research (NWO) Project no. 648.001.005, led by Carsten Dominik and Wim van Westrenen. TVH and AR acknowledge the financial support from the Belgian PRODEX program (grant no. 4000120791) managed by the European Space Agency in collaboration with the Belgian Federal Science Policy Office and from the BRAIN.be program of the Belgian Federal Science Policy Office. SC acknowledges financial support from OCAS NV by an OCAS-endowed chair at Ghent University. Work by GSN is supported by Deutsche Forschungsgemeinschaft (DFG, German Science Foundation) through Research Unit 2440 (Matter Under Planetary Interior Conditions, STE1105/13-1).


\appendix 

\section{Equation of State Extrapolations}\label{paper1:appendixExtrapolation}
\begin{figure*}[!ht]
  \centering	
    \includegraphics[width=.45\textwidth]{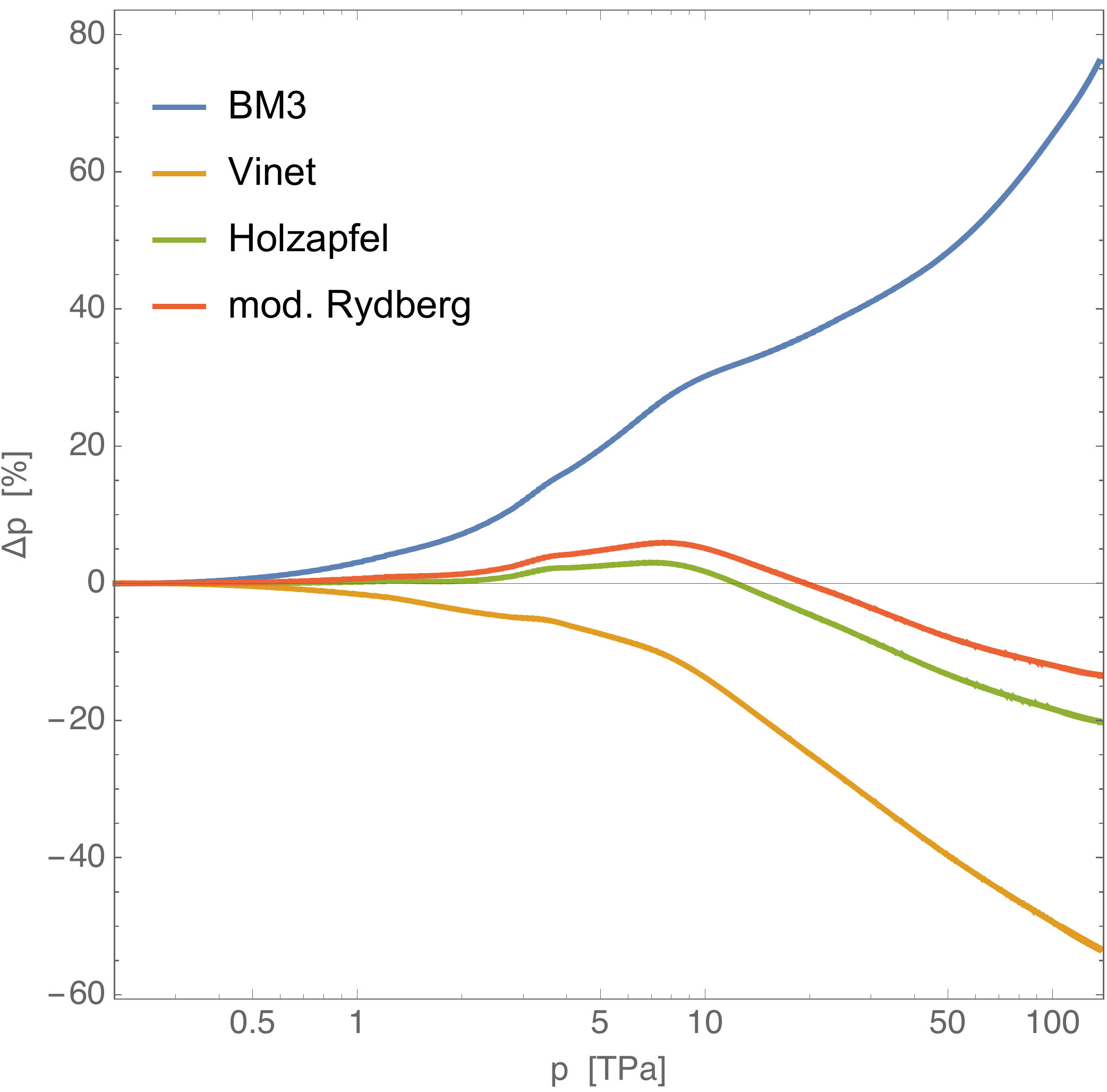}
	\includegraphics[width=.45\textwidth]{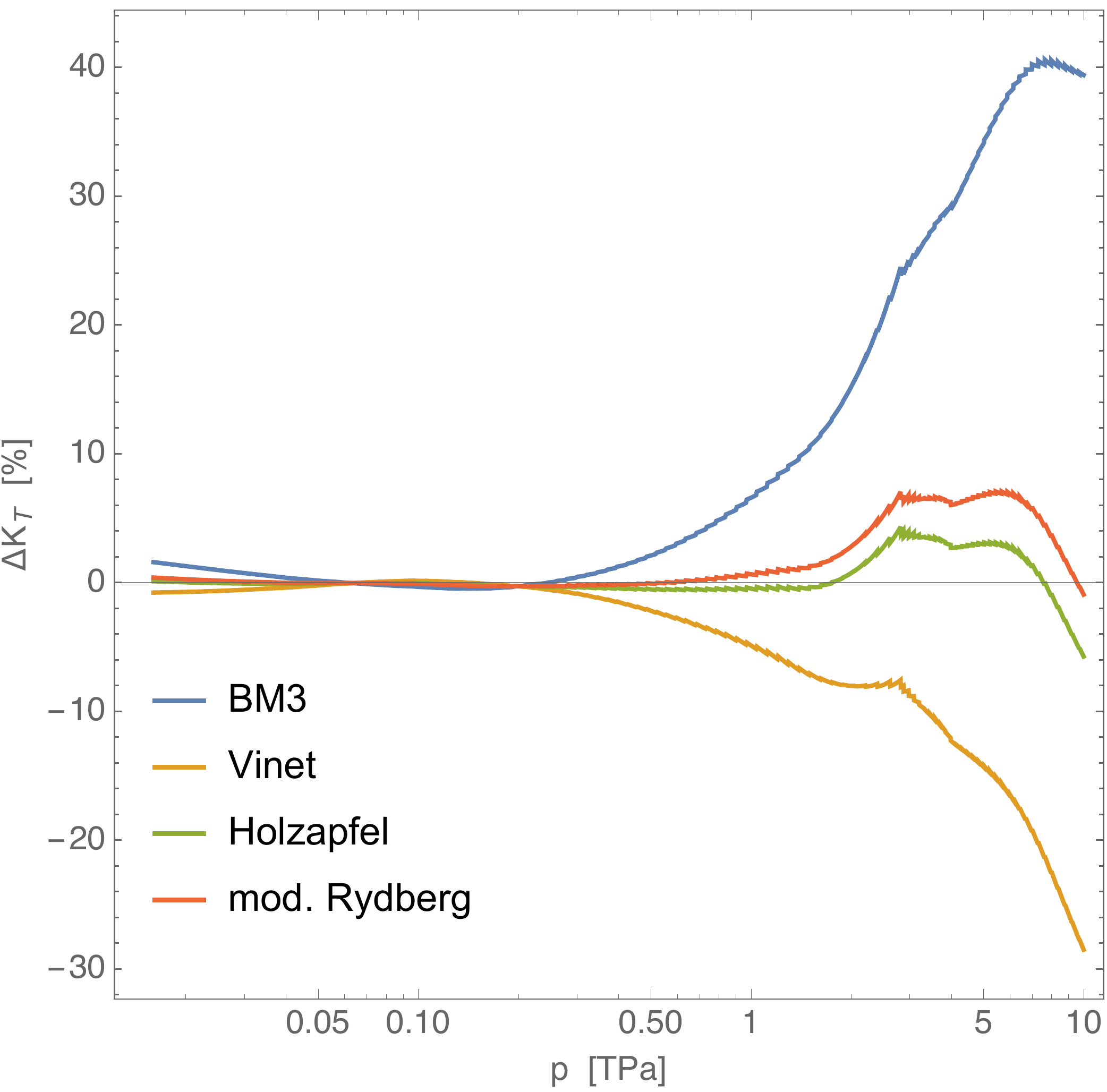}
  \caption{Relative differences in $P$ and $K_{T}$ between the DFT results and predictions obtained with the BM3, Vinet, modified Rydberg, and Holzapfel equation as a function of pressure.}
  \label{fig:deltapKT}
\end{figure*}

In order to assess which equation of state among the third-order Birch-Murnaghan (BM3), Vinet, modified Rydberg, and Holzapfel is more suitable for extrapolation, we fit them to the low pressure part (0$-$300~$\mathrm{GPa}$) of our DFT calculated $P(V)$ set (see Supplement) and compare the pressures predicted with our set of results (up to $10~\mathrm{TPa}$). For all equations, every parameter has been estimated without any prior constraints except for the modified Rydberg equation where $K_{\infty}'>5/3$ has been assumed. 

The BM3 and Vinet equations have residuals below $0.4~\mathrm{GPa}$ and those of the modified Rydberg and Holzapfel are below $0.05~\mathrm{GPa}$. This is significantly smaller than present day experimental error ({\textless} $2\%$) \citep{Fei2016}. Consequently, they are equally suitable for summarizing the low pressure data. Note that an unconstrained fit with the modified Rydberg equation has smaller residuals than the constrained fit, but the estimated value of $K_{\infty}'$ is $1.43$, smaller than the theoretical value of $5/3$ \citep{Stacey2005}.

Extrapolation with the Vinet and BM3 equation results in differences that are already above $2\%$ at pressures that are not much larger than $1~\mathrm{TPa}$ and above $14\%$ at $10~\mathrm{TPa}$ (Fig.~\ref{fig:deltapKT}). The pressures predicted by the modified Rydberg and Holzapfel equations differ significantly less from the DFT results (Fig.~\ref{fig:deltapKT}). Up to $10~\mathrm{TPa}$ the difference are below $3\%$ for the modified Rydberg (not shown) and Holzapfel equations. Differences with the 'constrained' modified Rydberg are about two times larger. Note that at $137~\mathrm{TPa}$ the predicted pressures with the BM3 and Vinet equation differ from the DFT set by more than $50\%$ and the modified Rydberg and Holzapfel equation by about $20\%$. For completeness we also compare $K_T$ predicted by the equations of state and directly calculated from the DFT set (Fig.~\ref{fig:deltapKT}).

\section{Fitting Residuals of DFT Data}\label{paper1:appendixFitting}
\begin{figure}[!h]
  \centering
    \includegraphics[width=.45\textwidth]{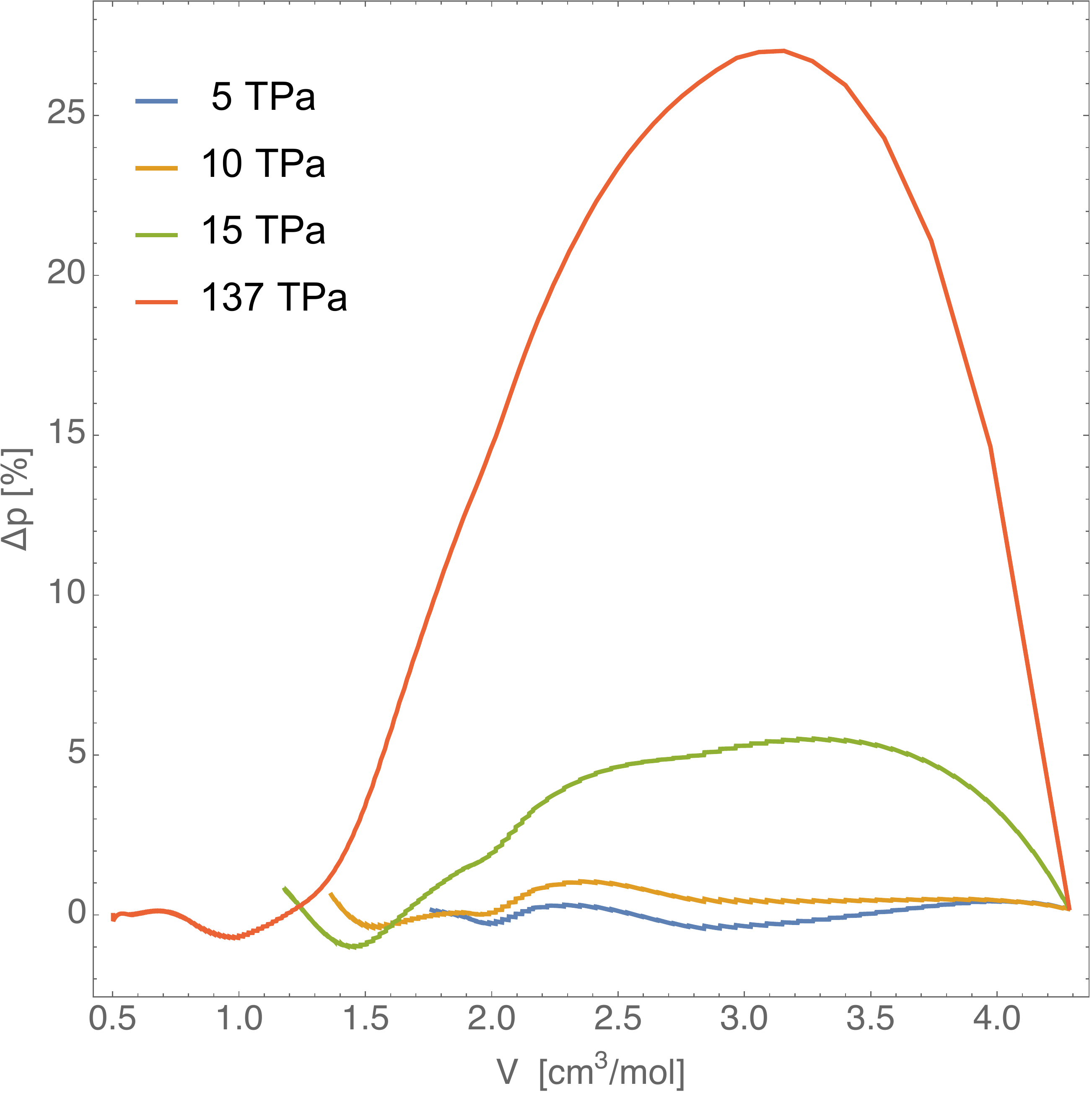}
  \caption{Fit residuals of the Holzapfel equation for different subsets of the DFT results.}
  \label{fig:deltap_V_Holzapfel}
\end{figure}
Fig.~\ref{fig:deltap_V_Holzapfel} demonstrates that the whole DFT result set (see Supplement) cannot be summarized accurately with one equation of state. However, a fit on a subset (0.234$-$10~$\mathrm{TPa}$) with the Holzapfel equation has residuals that are below $2\%$ if the reference volume is given (see main text). Comparable residuals are obtained for data at $P<5~\mathrm{TPa}$.

\section{Sliding Window Interpolation Scheme}\label{paper1:appendixInterpolation}
To compute volumes at intermediate pressures we use a sliding-window local third-order Birch-Murnaghan (BM3) fitting scheme. Let $\{V_i,E_i,P_i\}_{1\leq i\leq N}$ be the volume, energy, and pressure data set. The volume and bulk modulus at pressure $\tilde{P}$ is then calculated as follows:
\begin{enumerate}
	\item locate the position of the pressure in $\{P_i\}_{1\leq i\leq N}$ that is closest to $\tilde{P}$, say at $k$,
	\item select a small subset of length $2l+1$ of volumes and internal energies around $P_k$: $\{V_j,E_j\}_{k-l\leq j \leq k+l}$,
	\item define a local enthalpy as: $H(V)=E(V)+\tilde{P}\;V$ (per construction $H$ has a minima at $P=\tilde{P}$, since $\frac{d E}{dV}=-P$),
	\item estimate $V$ and $K_T$ at $\tilde{P}$ by fitting the BM3 equation on $\{V_j,H_j\}_{k-l \leq j \leq k+l}$.
\end{enumerate}
This scheme can be used to compute $V$ and $K_T$ on the fly or beforehand. In practice, the size of the local subset is chosen large enough to allow for a precise fit but not much larger. We have used $l=7$.

\section{Equation of State of Iron-alloys}\label{paper1:appendixFeAlloys}
\begin{table}[!h]
\caption{\label{tab:SataHolzapfelEOS} Thermoelastic parameters obtained by fitting the data of FeS, FeSi, Fe$_{0.95}$O and Fe$_{3}$C from \citet{Sata2010} to a Holzapfel equation. } 
\scriptsize
\begin{center}
\begin{tabular}{l|llll} \hline \hline \rule[0mm]{0mm}{0mm}
Fe-alloy  &$\rho_{0}$ & $K_{T,0}$ & $K'_{T,0}$ & $P_{FG0}$ \\

& ($\frac{\mathrm{kg}}{\mathrm{m^{3}}}$) & (GPa)& & (TPa) \\[2mm]																
\hline
FeS           & 6118 & 144.3 & 4.79 & 27.78  \\
FeSi          & 6522 & 219.5 & 4.29 & 18.45  \\
Fe$_{0.95}$O  & 5764 & 152.7 & 4.13 & 11.25  \\
Fe$_{3}$C     & 7980 & 289.4 & 3.81 & 30.68  \\[2mm]
\hline
\end{tabular}
\end{center}\caption*{ 
Key: $\rho_{0}$: reference density at $P_{0}=1$ bar and $T_{0}=300$~K, $K_{T,0}$: reference isothermal bulk modulus and its derivative $K'_{T,0}$, $P_{FG0}$: Fermi-gas pressure. }
\end{table}

\citet{Sata2010} determined the density of FeS, FeSi, Fe$_{0.95}$O and Fe$_{3}$C experimentally in a pressure range relevant for the Earth's core and fitted the data to a BM3 equation. In the absence of data in the high-pressure range of the super-Earth cores, extrapolations significantly beyond the experimental pressure range are needed. Such extrapolations can result in unreasonably large deviations (see Fig.~\ref{fig:EOSIronComparison}). Since extrapolation with a Holzapfel equation leads to better results than with a BM3 equation (Fig.~\ref{fig:deltap_V_Holzapfel}), we have refitted the density data from \citet{Sata2010} to a Holzapfel equation (Table \ref{tab:SataHolzapfelEOS}).

\section{DFT Equations of State of Iron }\label{paper1:appendixLiteratureEOS}
\begin{figure*}[!h]
  \centering
    \includegraphics[width=.45\textwidth]{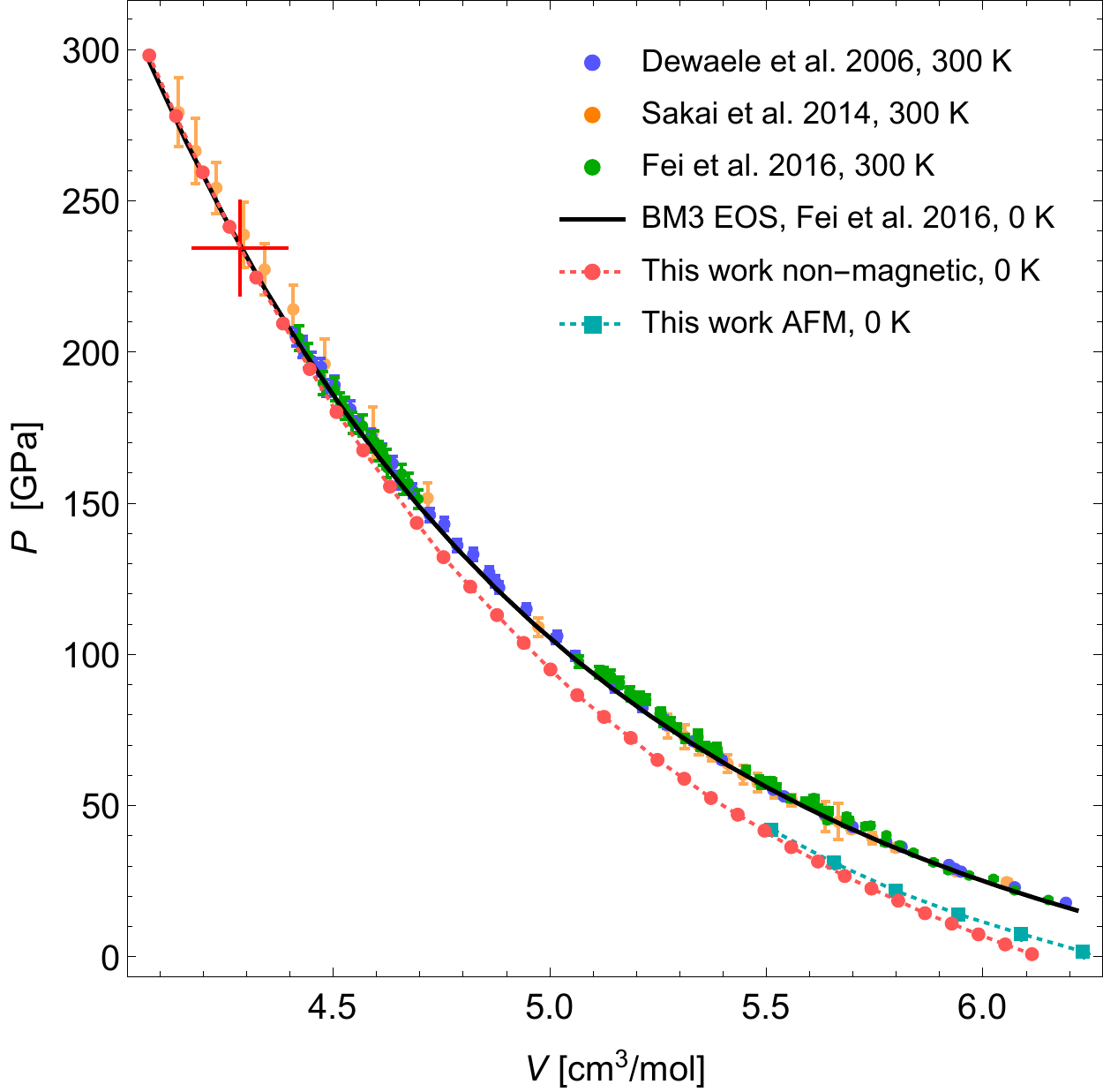}
  \caption{Pressure of hcp iron as a function of volume. Recent laboratory data with error-bars, DFT results for non-magnetic and anti-ferromagnetic phases, and the BM3 equation of state from \citet{Fei2016} corrected to 0~K and zero-point vibrational effects \citep{Lejaeghere2014}. The red plus denotes the pressure (234.4~GPa) and volume (4.28575~cm$^{3}$/mol) at which our DFT results match the experimental equation of state from \citet{Fei2016}.}
  \label{fig:LowPressureSEOS}
\end{figure*}

In Fig. \ref{fig:LowPressureSEOS}, we compare our DFT results at low pressures (0$-$300~GPa) with similar DFT predictions and laboratory data from the literature. We compare our DFT results at zero pressure with studies from the literature in Table \ref{tab:IronEOSTable}.

\begin{table*}[!h]
\caption{\label{tab:IronEOSTable} Ab initio equations of state of hcp-Fe. } 
\scriptsize
\begin{center}
\begin{tabular}{llll|l} \hline \hline \rule[0mm]{0mm}{0mm}
$V_{0}$ & $K_{0}$ & $K'_{0}$ & XC & Reference \\

($\mathrm{\frac{cm^{3}}{mol}}$) & (GPa)& & & \\[2mm]																
\hline
6.145           & 291       & 4.4       & PW91 & \citet{Stixrude1994}, NM \\ 
6.136$-$6.250   & 287$-$296 & 4.2$-$4.5 & PW91 & \citet{Alfe2000}, NM \\ 
6.175           & 293       & $-$       & PW91 & \citet{Caspersen2004}, NM \\  
\hline
6.158           & 292       & 4.4       & PBE  & \citet{SteinleNeumann1999}, NM \\ 
6.078           & 296       & $-$       & PBE  & \cite{ShaCohen2010}, NM \\ 
6.135           & 291       & 5.1       & PBE  & This Work, NM \\ 
\hline 
6.315           & 224       & 5.5       & PBE  & \citet{Dewaele2006}, AFM \\
6.354           & 209       & 5.2       & PBE  & \cite{SteinleNeumann1999}, AFM \\
6.306           & 222       & 5.7       & PBE  & This Work, AFM \\[2mm]
\hline
\end{tabular}
\end{center}\caption*{ Key: XC: Exchange correlation functional of hcp-Fe, NM: non-magnetic, AFM: antiferromagnetic, PW91: \citet{Perdew1992}, PBE: \citet{Perdew1996}. Reference temperature and pressure are zero. }
\end{table*}
  
\section{Equations of State of Iron and Effects on Kepler-36b}\label{paper1:appendixKepler36bEOS}
%
\begin{table*}[!h]
\caption{\label{tab:LiteratureEOSTable} Thermoelastic parameters for various equations of state of iron from the literature. }
\scriptsize
\begin{center}
\begin{tabular}{ll|cccc|cccc|cccc|l} \hline \hline \rule[0mm]{0mm}{0mm}
Ref. & Form.  & \multicolumn{4}{c}{Isothermal} & \multicolumn{4}{c}{Harmonic} & \multicolumn{4}{c}{Anharmonic + Electronic} & \\
& & $\rho_{0}$ & $K_{T,0}$ & $K'_{T,0}$ & $K'_{T,\infty}$ & $\Theta_{0}$ & $\gamma_{0}$ & $b$ [$q$] & $\gamma_{\infty}$ & $a_{0}$ & $m$ & $e_{0}$ & $g$ & $P$ range\\

& & ($\frac{\mathrm{kg}}{\mathrm{m^{3}}}$) & (GPa)& & & (K) & & & & ($\frac{1}{\mathrm{10^{3}~K}}$) & & ($\frac{1}{\mathrm{10^{3}~K}}$) &  & (GPa) \\[2mm]																
\hline

Dew & Vin & 8270 & 163.4  & 5.38  & $-$  & 417    & 1.875 &  3.289 & 1.305 & 0.037 & 1.87 & 0.195 & 1.339 & 17$-$197 \\ 
Bou & Hol & 8878 & 253.8  & 4.719 & $-$  & 44.574 & 1.408 & 0.826 & 0.827 & 0.2121 & 1.891 & $-$  & $-$  & 30$-$1500 \\
Val & Vin & 8300 & 160.2  & 5.82  & $-$  & 998    & 1.36  & [0.91]  & $-$  & $-$  & $-$  & $-$  & $-$  & 136$-$300 \\
Sea & Vin & 8300 & 156.2  & 6.08  & $-$  & $-$    & $-$   & $-$  & $-$  & $-$  & $-$  & $-$  & $-$  & 0$-$330 \\ 
Fei & BM3 & 8269 & 172.7  & 4.79  & $-$  & 422    & 1.74  & [0.78]  & $-$  & $-$  & $-$  & $-$  & $-$ & 18$-$280 \\
Dor & BM3 & 8341 & 173.98 & 5.297 & $-$  & $-$    & 2.434 & [0.489]  & $-$  & $-$  & $-$  & $-$  & $-$  & 18$-$360 \\
Wag & Ryd & 8269 & 149.4  & 5.65  & 2.94 & 430    & 1.875 &  3.289 & 1.305 & 0.037 & 1.87 & 0.195 & 1.339 & 17$-$197 \\[2mm]
\hline
\end{tabular}
\end{center}\caption*{ Key: $\rho_{0}$: reference density at ambient conditions, $K_{0}$: reference isothermal bulk modulus and its derivatives $K'_{T,0}$ and $K'_{T,\infty}$. Note: \citet{Fei2016} uses a different formulation for the electronic contribution to the specific heat, $C_{Ve}=\beta_{0}x^{3k}T$, where $\beta_{0}=0.07$~J kg$^{-1}$ K$^{-2}$ and $k=1.34$, and $\gamma_{e}=2$. \\ Abbreviations: Dew: \citet{Dewaele2006}; Bou: \citet{Bouchet2013}; Val: \citet{Valencia2007a}, data from \citet{Williams1997}; Sea: \citet{Seager2007}, data from \citet{Anderson2001}; Fei: \citet{Fei2016}; Dor: \citet{Dorn2015}, data from \citet{Belonoshko2010}; \citet{Wagner2011}, data from \citet{Dewaele2006}; Vin: Vinet; BM3: third-order Birch-Murnaghan; Hol: Holzapfel; Ryd: modified Rydberg. }
\end{table*}

\begin{figure*}[!h]
  \centering
  \medskip
  \includegraphics[width=0.9\textwidth]{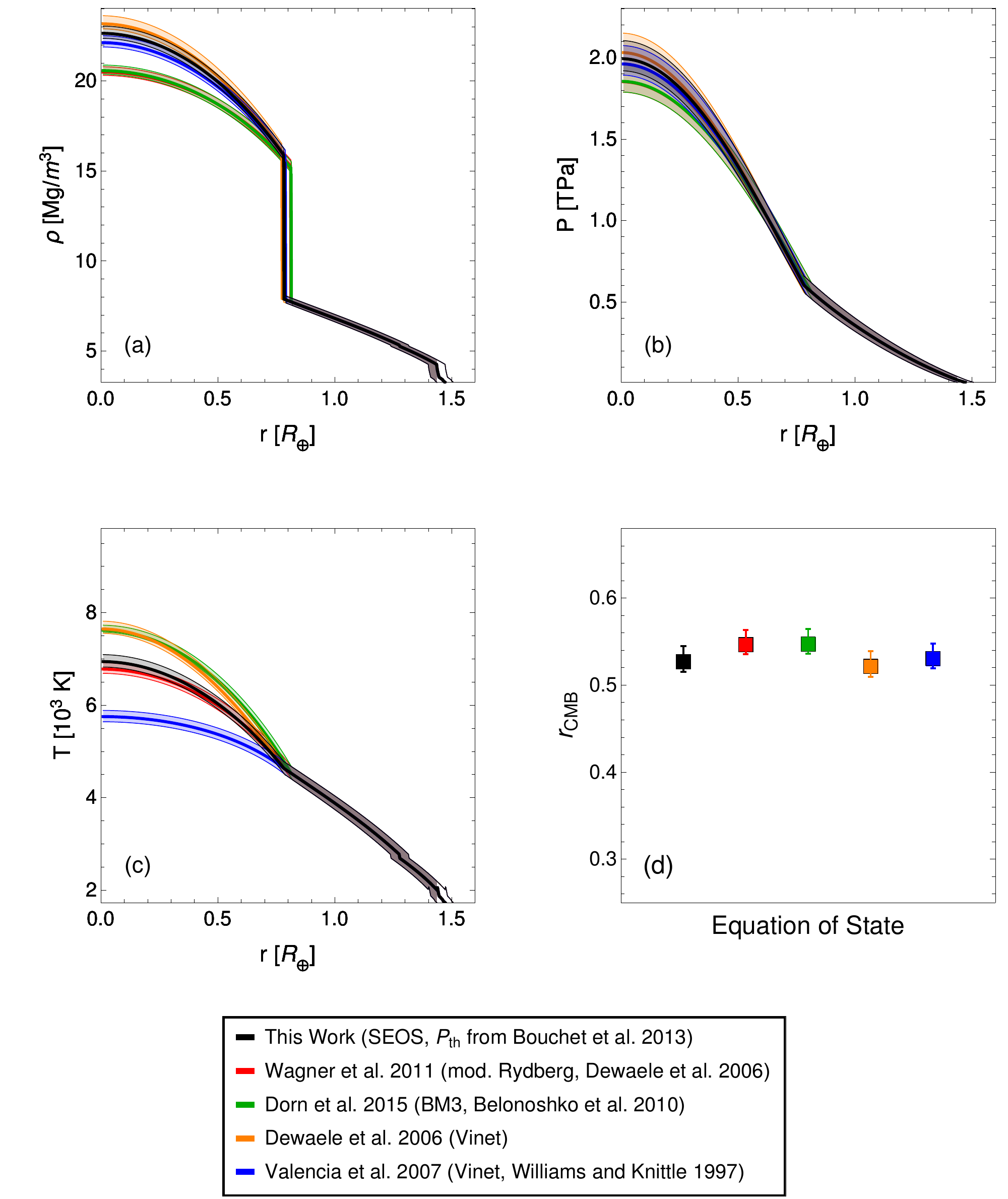}
  \caption[The interior structure of Kepler-36b for different equation of state]{Density, pressure and temperature distribution, and the core radius fraction of Kepler-36b calculated using different equations of state of iron. Shaded areas represent uncertainties arising due to uncertainties in mass and radius of Kepler-36b. }
  \label{fig:Kepler36bPlotEOS}
\end{figure*}

Table \ref{tab:LiteratureEOSTable} lists the thermoelastic parameters of various equations of state of iron from the literature used for modeling planetary cores. In Fig.~\ref{fig:Kepler36bPlotEOS}, we illustrate the differences between our DFT-based equation of state and some equations of state from Table \ref{tab:LiteratureEOSTable} on the internal properties of Kepler-36b.  

\afterpage{
\clearpage
}

\afterpage{
\clearpage
}






\bibliographystyle{icarus}
\bibliography{SuperEarthsEOS}







\end{document}